\documentclass[twocolumn]{aastex61}
\pdfoutput=1
\usepackage{amsmath,amstext}
\usepackage[T1]{fontenc}
\usepackage[figure,figure*]{hypcap}
\usepackage{tikz}
\usetikzlibrary{shapes,arrows}

\newcommand{\logg}{$\log{g}$ }
\newcommand{\loghhe}{$\log{\mathrm{H}/\mathrm{He}}$ }
\newcommand{\logche}{$\log{\mathrm{C}/\mathrm{He}}$ }
\newcommand{\logcahe}{$\log{\mathrm{Ca}/\mathrm{He}}$}
\newcommand{\teff}{$T_\mathrm{eff}$ }
\newcommand{\msun}{$ M_\odot$}

\newcommand{\che} {$\log\ ({\rm C/He})$}

\newcommand{\halpha} {H$\alpha$ }

\newcommand{\Te} {$T_{\rm eff}~$}

\shorttitle{Analysis of Polluted Helium-Rich White Dwarfs}
\shortauthors{Coutu et al.}

\begin{document}

\title{Analysis of Helium-Rich White Dwarfs Polluted by Heavy Elements
  in the\\Gaia Era}

\author{S. Coutu}
\affiliation{D\'epartement de Physique, Universit\'e de Montr\'eal, Montr\'eal, 
QC H3C 3J7, Canada; coutu@astro.umontreal.ca, dufourpa@astro.umontreal.ca.,bergeron@astro.umontreal.ca,sblouin@astro.umontreal.ca}
\author{P. Dufour}
\affiliation{D\'epartement de Physique, Universit\'e de Montr\'eal, Montr\'eal, 
QC H3C 3J7, Canada; coutu@astro.umontreal.ca, dufourpa@astro.umontreal.ca.,bergeron@astro.umontreal.ca,sblouin@astro.umontreal.ca}
\author{P. Bergeron}
\affiliation{D\'epartement de Physique, Universit\'e de Montr\'eal, Montr\'eal, 
QC H3C 3J7, Canada; coutu@astro.umontreal.ca, dufourpa@astro.umontreal.ca.,bergeron@astro.umontreal.ca,sblouin@astro.umontreal.ca}
\author{S. Blouin}
\affiliation{D\'epartement de Physique, Universit\'e de Montr\'eal, Montr\'eal, 
QC H3C 3J7, Canada; coutu@astro.umontreal.ca, dufourpa@astro.umontreal.ca.,bergeron@astro.umontreal.ca,sblouin@astro.umontreal.ca}
\author{E. Loranger}
\affiliation{D\'epartement de Physique, Universit\'e de Montr\'eal, Montr\'eal, 
QC H3C 3J7, Canada; coutu@astro.umontreal.ca, dufourpa@astro.umontreal.ca.,bergeron@astro.umontreal.ca,sblouin@astro.umontreal.ca}
\author{N.F. Allard}
\affiliation{GEPI, Observatoire de Paris, Universit\'e PSL, CNRS, UMR 8111,
  61 avenue de l'Observatoire, F-75014 Paris, France}
\affiliation{Sorbonne Universit\'e, CNRS, UMR 7095,
  Institut d'Astrophysique de Paris, 98bis boulevard Arago, F-75014 Paris, France}
\author{B.~H. Dunlap}
\affiliation{Department of Physics and Astronomy, University of North
Carolina at Chapel Hill, Chapel Hill, NC 27599, USA}
\affiliation{Department of Astronomy and McDonald Observatory, University of Texas at Austin, Austin, TX 78712, USA}
\affiliation{McDonald Observatory, Fort Davis, TX-79734, USA}

\begin{abstract}
We present a homogeneous analysis of 1023 DBZ/DZ(A) and 319 DQ white
dwarf stars taken from the Montreal White Dwarf Database. This
represents a significant increase over the previous comprehensive
studies on these types of objects. We use new trigonometric parallax
measurements from the {\it Gaia} second data release, together with
photometry from the Sloan Digital Sky Survey, Pan-STARRS, {\it Gaia},
or \textit{BVRI} from the literature, which allow the determination of
the mass for the majority of the objects in our sample. We use the
photometric and spectroscopic techniques with the most recent model
atmospheres available, which include high-density effects, to
accurately determine the effective temperature, surface gravity, and
heavy element abundances for each object. We study the abundance of
hydrogen in DBZ/DZ white dwarfs and the properties of the accreted
planetesimals. We explore the nature of the second sequence of DQ
stars using proper motions from {\it Gaia}, and highlight evidence of
crystallization in massive DQ stars. We also present mass
distributions for both spectral types. Finally, we discuss the
implications of our findings in the context of the spectral evolution
of white dwarfs, and provide the atmospheric parameters for each star.
\end{abstract}

\keywords{stars: abundances  -- stars: atmospheres -- stars: evolution -- white dwarfs}

\section{Introduction}
 
White dwarfs represent the final evolutionary phase of main sequence
stars with initial mass below $\sim$8 \msun~ and are characterized by
their high surface gravity, typically $g=10^8$ cm s$^{-2}$. Because of
this, elements heavier than helium will sink below the photosphere in
characteristic timescales that are many orders of magnitude smaller
than the cooling age of the star \citep{Paquette1986}. This
gravitational separation also explains why most white dwarfs --- about
80\% --- have pure hydrogen atmospheres. The only absorption lines
present in their spectra are those of hydrogen, and they are
collectively known as DA stars. For a smaller fraction, practically no
hydrogen survives the late phases of stellar evolution, and a thin
opaque helium layer --- the heaviest element remaining --- will float
on top and form the atmosphere. Depending on the effective
temperature, they are classified as DO white dwarfs if they show
ionized helium lines, and DB stars if only neutral helium lines can be
observed. Below $T_{\rm eff}\sim 12,000$~K, spectra of pure helium
atmosphere white dwarfs become featureless as there is not enough
energy to populate the lower energy levels of He \textsc{i} line
transitions. Such objects with continuous optical spectra are
classified as DC white dwarfs. Note that a similar phenomenon also
happens for hydrogen-rich white dwarfs cooler than about 5000~K as the
electrons are mostly found in the ground state, preventing Balmer line
transitions. In the absence of physical mechanisms competing with
gravitational settling, the optical spectra of all white dwarfs should
thus show only hydrogen lines, helium lines, or pure
continuum. Nevertheless, white dwarfs with traces of heavy elements do
exist, indicating that gravitational settling is not acting
alone. These ``contaminated'' white dwarfs are found mainly in two
categories: i) those with traces of heavy elements (other than carbon)
in their optical spectra, and ii) those with primarily carbon
absorption lines, either molecular or atomic.

\subsection{Heavy Element Pollution}

White dwarfs showing absorption lines from elements such as calcium,
magnesium, or iron are collectively known as DZ stars if their spectra
show only heavy element lines, DBZ stars if they show helium and heavy
elements, and DAZ stars if they display hydrogen and metal
features. Model atmospheres show that both the DZ and DBZ stars are
helium dominated while the DAZs are hydrogen dominated. The presence of
heavy elements in these objects is now understood as external
accretion of matter from a disk of debris resulting from the
destruction of a rocky object (asteroid or small planets) by the white
dwarf's tidal forces \citep[see][and references
  therein]{Jura2014}. These objects are thus temporarily imprinted
(diffusion timescales range from a few hundred thousands to a few
million years\footnote{Based on calculations by G.~Fontaine, included
  in the MWDD and available at
  \url{http://www.montrealwhitedwarfdatabase.org/evolution.html}},
much shorter than the cooling age) with the chemical composition of
the polluting body, providing a unique opportunity to study the
chemical composition of extra-solar bodies
\citep{Zuckerman2007,Klein2010,Klein2011, Dufour2012, Jura2012,
  Xu2014,Koester2014, Xu2016, Xu2017, Blouin2019}. High-resolution or
UV spectroscopic observations of some samples have shown that 25 to
50\% of all white dwarfs are contaminated by heavy elements at some
level \citep{Zuckerman2003, Zuckerman2010, Koester2014}.

Moreover, since observational evidence indicates that these white
dwarfs had (or still have) at least some sort of planetary system
around them, they can provide information about the correlation
between stellar mass and planet occurrence, which can give insight
into the planet formation process \citep{Johnson2010}. Indeed, most
observational exoplanet surveys are biased towards lower mass
stars. The Kepler mission prioritized G-type stars
\citep{Batalha2010}, which have masses around 1 $M_\odot$. More
massive stars of type O and B are observed much less frequently by
Kepler, mostly because of their scarcity (they spend a very short time
on the main sequence), and the Kepler field of view was chosen to
avoid young stellar populations \citep{Batalha2010}. Doppler surveys
also favor Sun-like stars, as their spectral properties make their
detection easier \citep{Johnson2010}. Thus, very little is known about
the relation between planet occurrence and stellar mass above
$M=2-3\ M_\odot$. While massive white dwarfs --- which had massive
main sequence progenitors --- are fainter, the selection bias is much
less important, especially with {\it Gaia} DR2, which is expected to
be volume-complete within $\sim$70 parsecs
\citep{Gentile2019}. Inferences of planetary systems around white
dwarfs are thus not subjected to the same limitations.

\subsection{Carbon Pollution}

White dwarfs showing mainly carbon features are collectively known as
DQ stars. They represent 9\% of the white dwarfs in the local sample
\citep[$D < 20$ pc,][]{Giammichele2012}. They show only atomic carbon
lines when $T_{\rm eff}\gtrsim 10,000$ K, and molecular C$_2$ bands at
lower effective temperatures, with a smooth transition around that
temperature where both molecular bands and atomic carbon lines are
simultaneously present. Model atmosphere analyses have shown that they
are helium-dominated with carbon abundances ranging from \che = $-$7
to $-$2 \citep{Weidemann1995, Dufour2005, KoesterKnist}, and effective
temperatures between $\sim$5000 K and 12,000 K. A model where the deep
helium convection zone catches up with the settling carbon in the
core, and dredges it up to the surface, can successfully account for
the observed abundances in most objects \citep{Fontaine2005,
  Dufour2005}. \citet{Dufour2005} showed, however, that several DQ
stars had larger than average carbon abundances, forming a distinct
sequence about 1 dex above the bulk of the sample in a \che~vs
\Te\ diagram \citep[see also][]{KoesterKnist}. Since the only object
with a measured trigonometric parallax belonging to this second
sequence was massive, Dufour et al.~proposed that they could represent
the high-mass tail of the white dwarf mass distribution. However,
\citet{brassard2007} showed that an evolutionary sequence at 1
\msun~does not correctly predict the carbon abundance pattern that is
empirically observed, indicating that another explanation to account
for these stars must be sought.

While the cooler end of the DQ sequence is consistent with the
expectations from the dredge-up model, very few stars were known at
that time to test the theory on the hotter side (\Te $\ge$
13,000~K). Many hot objects showing mainly ionized or neutral carbon
lines (and also oxygen in a few cases) had been identified in the
Sloan Digital Sky Survey (SDSS) by \citet{Liebert2003}. The
interpretation then was that these objects were a hotter version of
the cool DQ white dwarfs, which had helium-dominated atmospheres with
traces of carbon. However, when the hottest objects were analyzed
using state-of-the-art model atmospheres, it was found in fact that
the main atmospheric constituent was carbon, not helium
\citep{Dufour2007Nat,Dufour2008}. \citet{Dufour2013} then showed that,
in a $u-g$ vs $g-r$ color-color diagram, the hot carbon-dominated
atmosphere DQ stars (\Te$\sim 18,000-24,000$~K with mainly ionized
carbon lines) seem to form a sequence that connects with warm DQ white
dwarfs with neutral atomic lines (\Te$\sim 10,000-16,000$~K), followed
by cooler DQ stars with strong molecular bands (\Te < 10,000~K on
the second sequence mentioned above). Unfortunately, very few
trigonometric parallax measurements were available for these objects
at that time, and thus the massive nature of the elusive sequence was
somewhat speculative.

Since the studies of \cite{Dufour2005,Dufour2007} --- the latest
comprehensive analyses of large samples of DQ and DZ stars,
respectively --- there have been several improvements in stellar
atmosphere modeling (for example, \citealt{Blouin2017},
\citealt{Blouin2018a}). Simultaneously, several surveys have enlarged
the sample of spectroscopically confirmed white dwarfs considerably,
and consequently, the number of known DQ/DZ/DBZ white dwarfs has
increased by more than a factor of five. Also, thanks to the second
{\it Gaia} data release in April 2018, distances are now available,
for the first time, for most of these objects, a quantity required to
obtain precise measurements of their stellar masses.

The availability of improved model atmospheres and new data motivated
us to perform an updated homogeneous analysis of all these
metal-polluted white dwarfs. We describe the model atmospheres in
Section \ref{Section:models}, the observational data in Section
\ref{Section:observations}, and the methodology in Section
\ref{section:method}. We present the analysis for the DBZ/DZ stars in
Section \ref{Section:DZ}, and for the DQ stars in Section
\ref{Section:DQ}. We then discuss the implications of our results on
our understanding of the spectral evolution of white dwarfs in Section
\ref{Section:spectralevolution}. Our conclusions follow in Section
\ref{Section:conclu}.

\section{Model Atmospheres} \label{Section:models}

Our DZ/DBZ/DQ model atmosphere code is similar to that outlined in
\citet{Dufour2005,Dufour2007}, but with several physical improvements
described at length in \citet{Blouin2018a, Blouin2018b}. Of particular
importance for the study of DZ/DBZ stars are the new line profile
calculations following the unified line shape theory of
\citet{Allard1999} for strong transitions, the most important being Ca
\textsc{ii} H\&K, Mg \textsc{i} $\lambda$2852 and Mg \textsc{ii}
$\lambda$2795/2802, the Mgb triplet, and Ca \textsc{i}
$\lambda$4226. Less important transitions use Lorentzian or
quasistatic van der Waals broadening profiles \citep[][D.~Koester,
  private communication]{Walkup1984}. For the DQ model atmospheres,
one of the main improvements over \cite{Dufour2005} is the replacement
of the ``just overlapping line approximation'' \citep{Zeidler1982}, to
describe the C$_2$ Swan band opacity, with a complete linelist
provided by J.~O.~Hornkohl (private communication; see
\citealt{Parigger2015} for details of the methodology). We find that
the use of this new linelist provides a much better representation of
the shape of the observed Swan bands in DQ white dwarfs, particularly
in the region around 4300 \AA, which was poorly fitted in
\citet{Dufour2005}, compared to the prescription of
\citet{Zeidler1982}, \citet{Brooke2013}, or Kurucz
linelists\footnote{\url{http://kurucz.harvard.edu/linelists.html}}. The
atomic linelist of Kurucz has also been replaced by the compilation
from the Vienna Atomic Line Database \citep[VALD,][]{Piskunov1995}.

For the DZ/DBZ(A) white dwarfs, we generated a 4-dimensional grid of
model atmospheres and synthetic spectra with $T_{\rm eff}$ varying
from 4000 K to 16,000 K by steps of 500 K, \logg from 7.0 to 9.0 by
steps of 0.5 dex, \logcahe\ from $-12$ to $-7$ by steps of 0.5 dex,
and \loghhe from $-7$ to $-3$ by steps of 1 dex. We also generated a
grid with no hydrogen. Metal-to-metal ratios have been fixed to that
of chondrites \citep{Lodders2003} with respect to calcium. Calcium
thus serves as a proxy for the abundances of all other heavy
elements. While metal-to-metal ratios certainly differ from that of
chondrites in many objects, our assumption provides a good first order
approximation of the contribution of free electrons and opacity from
heavy elements \citep[see][and also discussion below]{Dufour2007}.

For the DQ model grid, since no DQ white dwarfs are found at high
effective temperature and low carbon abundance, or at low effective
temperature and high carbon abundance, we generated two separate
3-dimensional grids. The first grid is generated with $T_{\rm eff}$
from 8000 K to 16,000 K by steps of 500 K, \logg from 7 to 9 by steps
of 0.5 dex, and \logche from $-5$ to $-1$ by steps of 0.5 dex. The
second grid covers $T_{\rm eff}$ from 6000 K to 10,000 K by steps of
500 K, \logg from 7 to 9 by steps of 0.5 dex, and \logche from $-8$ to
$-4$ by steps of 0.5 dex. No hydrogen is included in these models, but
some smaller grids with hydrogen were generated to test its effects,
as discussed in Section \ref{section:dqhydro}.

\section{Observations} \label{Section:observations}

\subsection{Note on Naming Convention}

Some white dwarfs can have up to 20 different names. Here we decided
to adopt names based on ICRS coordinates at epoch and equinox 2000
instead of mixing names from different catalogs. Stars will be named
JHHMM$\pm$DDMM, where the first four digits correspond to the right
ascension in hours and minutes, and the last four digits to the
declination in degrees and minutes in sexagesimal notation. In some
cases, a second relevant name will be written in parentheses. Table
\ref{tab:names} provides cross-references for the names of the objects
in our sample with {\it Gaia} source id, MWDD id, and the full
coordinates.

\subsection{Sample Selection}

We first selected from the Montreal White Dwarf
Database\footnote{\url{www.montrealwhitedwarfdatabase.org}}
\citep[][hereafter MWDD]{Dufour2017} all objects with either a Q or a
Z in their spectral classification. This includes DZ, DZA, DBZ, DBZA,
DQ, etc. We did not include DAZ white dwarfs because they have
hydrogen-rich atmospheres, and we are only interested here in
helium-rich atmospheres. Most objects in our sample have previously
been identified in various white dwarf catalogs from the Sloan Digital
Sky Survey (SDSS;
\citealt{Kepler2015,Kepler2016,Kleinman2004,Kleinman2013,Eisenstein2006}). Many
objects have also been classified as uncertain, such as DQ: or
DC-DQ. We thus visually inspected every spectrum and rejected those
that did not show the appropriate spectral lines, or were too noisy to
make a reliable classification (spectra with better signal-to-noise
will eventually confirm their spectral types). We rejected known
unresolved binary systems as well as magnetic white dwarfs. In order
to ensure a homogeneous analysis based exclusively on optical
features, we also rejected some objects classified as DQ white dwarfs
but with carbon features detected only in the ultraviolet. We also
excluded stars that had carbon-dominated atmospheres, the so-called
Hot DQ stars ($T_\textrm{eff} \gtrsim 18,000$ K), or any object for
which our preliminary analysis gave atmospheric parameters outside of
our model grid. Finally, DQpec white dwarfs with strong distorted Swan
bands are also left out of our analysis because there are still large
uncertainties regarding their modeling \cite[see][]{Blouin2019c}.

All objects in our sample were then cross-matched with {\it Gaia} DR2
to retrieve photometric and astrometric data. For one object,
J0739+0513 (Procyon B), we use the parallax from Hipparcos
\citep[$284.56 \pm 1.26$,][]{vanleeuwen2007}. A few objects with
negative parallaxes or very large uncertainties ($\sigma_\pi/\pi > 1$)
were rejected; these are unlikely to be white dwarfs, and if they are,
the data are too imprecise to allow any satisfactory
analysis. \cite{bailerjones2015} demonstrated that inverting the
parallax to estimate the distance might lead to unreliable distances
when the uncertainty on the parallax is larger than 20\%, and that it
can lead to incorrect error estimates. Fortunately, 87\% of our
parallax sample have smaller uncertainties. In fact, the difference
between a probabilistic analysis distance and the inverse of the
parallax is less than 1\% for the majority of objects when $\sigma_\pi
/ \pi < 0.1$.

\begin{figure}
    \centering
    \includegraphics[width=0.7\columnwidth]{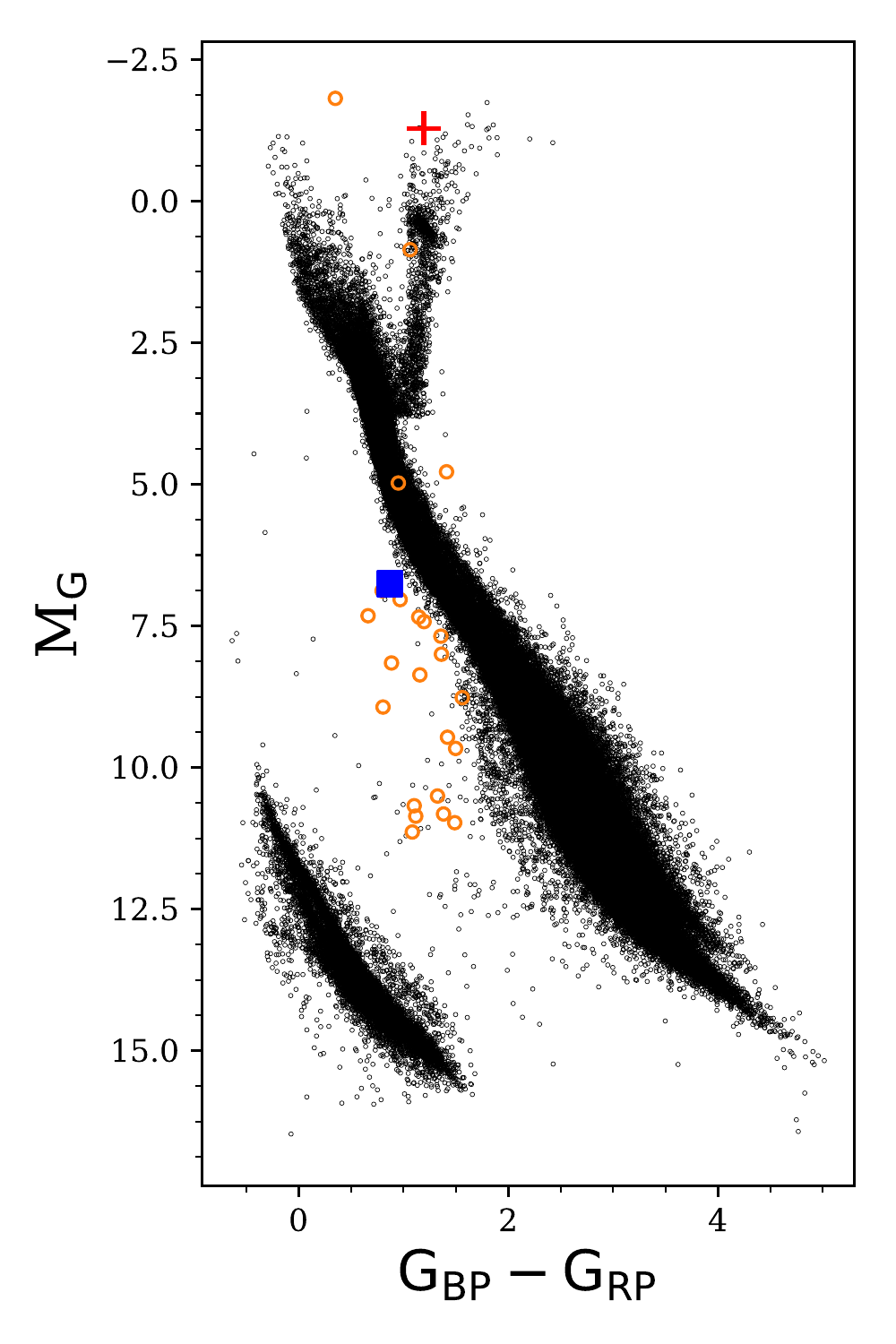}
    \caption{{\it Gaia} HR diagram for objects misclassified as white
      dwarfs (orange circles). SDSS J174618.94+262217.0 and SDSS
      J192013.71+383917.7, the two examples discussed in the text and
      displayed in Figure \ref{fig:wrongtypes}, are represented by a
      red cross and a blue square, respectively. Black dots are {\it
        Gaia} objects within 100 pc selected using the cuts proposed
      in \citet[][Appendix B]{Gaia2018}. White dwarf candidates are
      located in the bottom left portion of the diagram.}
  \label{fig:wrongtypes_colmag}
\end{figure}

\begin{figure}
	\centering
    \includegraphics[width=\columnwidth]{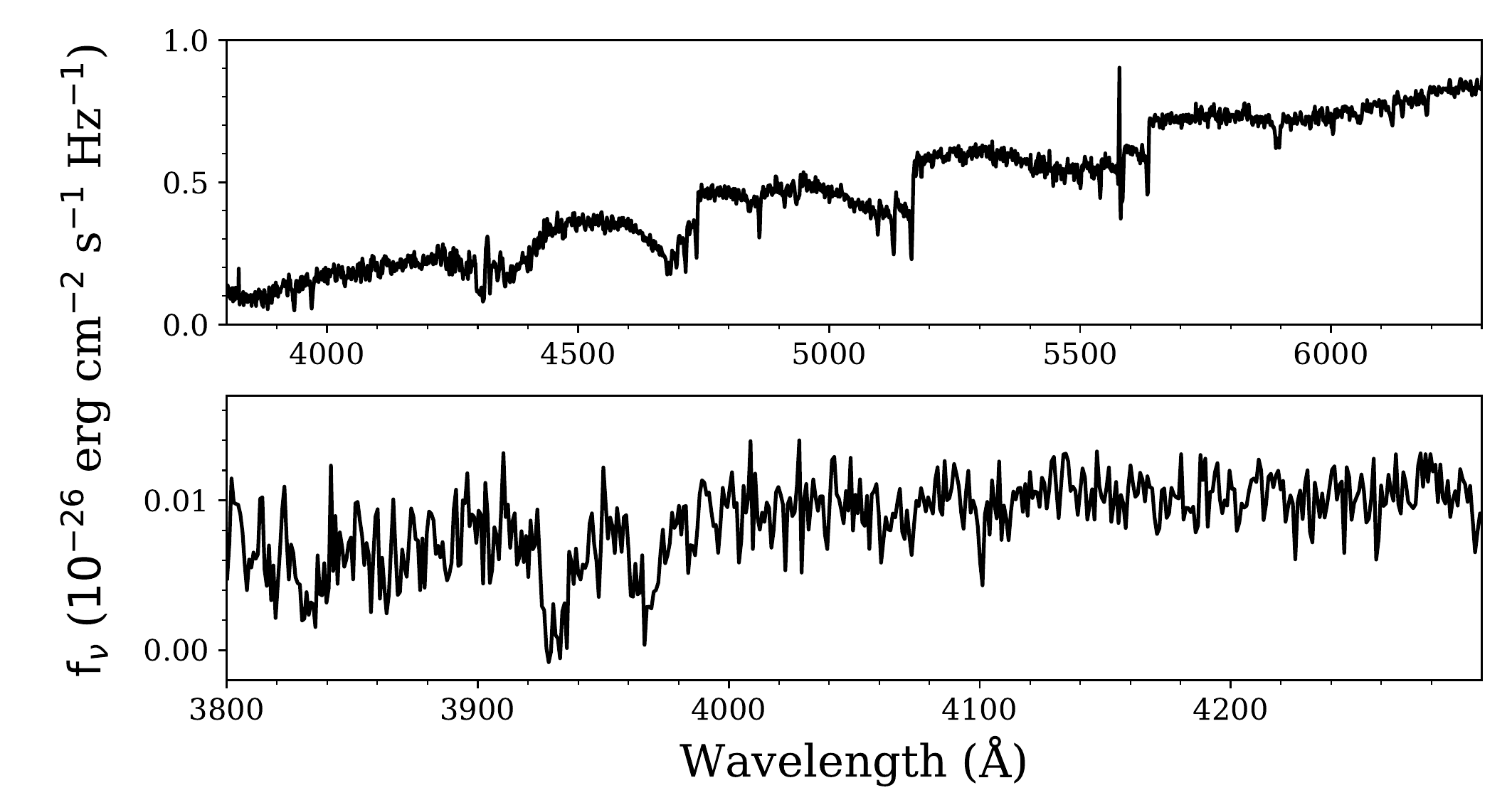}
    \caption{Optical spectra of SDSS J174618.94+262217.0 (top) and
      SDSS J192013.71+383917.7 (bottom), two objects that were
      previously misclassified as DQ and DZ white dwarfs,
      respectively.}
  \label{fig:wrongtypes}
\end{figure}

With the availability of trigonometric parallaxes and broadband colors
from {\it Gaia} for the majority of the objects in our sample, it is
now possible to place them accurately in an observational
Hertzsprung-Russell (HR) diagram \citep{Gaia2018}. This is illustrated
in Figure \ref{fig:wrongtypes_colmag}, where stars that have been
erroneously classified as white dwarfs can easily be
identified. Examples of such stars are the giant star SDSS
J174618.94+262217.0 \citep{green2013} and SDSS J192013.71+383917.7,
two objects that have been erroneously classified as a DQ and a DZ
white dwarf, respectively \citep{Kleinman2013}. The similarity of the
spectra of these non-degenerate stars, displayed in Figure
\ref{fig:wrongtypes}, with genuine DQ and DZ white dwarfs stresses the
importance of using photometric and parallax measurements in
combination to spectroscopic data during the classification
process. Overall, 25 objects, listed in Table \ref{tab:rejected},
previously classified as white dwarfs in various catalogs were
rejected this way.

\begin{table*}
\small
\centering
\caption{Objects rejected due to their location in the {\it Gaia} HR diagram}\begin{tabular}{ccc}
\hline
J Name & Gaia source id & MWDD id \\
\hline
J0205+0058 & 2508108382280944768 & 2MASS J02053812+0058354 \\
J2112+0912 & 1743430270303257728 & 2MASS J21120843+0912019 \\
J1642+3617 & 1328051961501737088 & Cl* NGC 6205 KAD 606 \\
J0239$-$0003 & 2498734800840646400 & SDSS J023955.79$-$000312.0 \\
J0814+3508 & 902942352007933568 & SDSS J081424.51+350822.9 \\
J0815+1537 & 655412730925191808 & SDSS J081502.24+153709.7 \\
J0914$-$0022 & 3842424562164246656 & SDSS J091455.49$-$002219.2 \\
J1136+4702 & 787028812551381248 & SDSS J113612.16+470206.8 \\
J1208$-$0008 & 3698751446481826432 & SDSS J120853.36$-$000847.3 \\
J1228+2057 & 3952475922233611776 & SDSS J122829.50+205747.0 \\
J1336+1535 & 3742286441879709952 & SDSS J133644.21+153554.2 \\
J1357+1252 & 3727838511198756224 & SDSS J135713.37+125230.9 \\
J1507+1824 & 1211873229279489536 & SDSS J150741.36+182406.8 \\
J1549+0436 & 4426232766560567296 & SDSS J154918.87+043651.0 \\
J1610+0247 & 4412254439017190400 & SDSS J161041.62+024706.2 \\
J1714+2324 & 4568671714800327680 & SDSS J171432.25+232423.0 \\
J1746+2622 & 4582265290590354304 & SDSS J174618.94+262217.0 \\
J1920+3839 & 2052876071205883520 & SDSS J192013.71+383917.7 \\
J2105+0828 & 1744023628624108288 & SDSS J210511.38+082853.7 \\
J2106+0642 & 1736767710874927872 & SDSS J210632.52+064233.5 \\
J2111+0915 & 1743431683347518464 & SDSS J211157.26+091554.2 \\
J2153+2750 & 1800030173963458048 & SDSS J215306.03+275057.3 \\
J2259$-$1001 & 2606526522781415168 & SDSS J225931.60$-$100146.9 \\
J0124+0541 & 2563653437678125568 & USNO-B1.0 0956$-$00013686 \\
J0257+0111 & 339405495907200 & WD 0255+009.1 \\
\hline
\end{tabular}
\label{tab:rejected}
\end{table*}

To constitute our final sample, we removed 63 stars that had
parameters outside our model grids (these will be analyzed elsewhere)
and 63 white dwarfs for which no optical spectroscopy was
available. In the end, we are left with a sample of 1023 DZ (679 with
parallax measurements) and 317 DQ (303 with parallax measurements)
white dwarfs, which we analyze in a homogeneous fashion in the next
sections.

\subsection{Photometric and Spectroscopic Data} \label{section:photo}

We cross-matched our sample with Pan-STARRS DR1 and SDSS DR14 to
retrieve $grizy$ and $ugriz$ PSF magnitudes, respectively. Since SDSS
photometry is available for 95\% of our sample, while only 80\% of our
objects have Pan-STARRS photometry, we decided to rely primarily on
SDSS photometry for the sake of homogeneity. Note that although
Pan-STARRS covers a much larger part of the sky than SDSS, most
spectroscopically identified white dwarfs were discovered in the
SDSS. Furthermore, SDSS has proven to be reliable many times
\citep[see][]{GenestBeaulieu2019}, while possible Pan-STARRS biases
have not been fully explored yet. When SDSS $ugriz$ is not available
for an object, we rely on Pan-STARRS data, previously published $BVRI$
photometry, or simply {\it Gaia} broadband photometry, in that order
of priority. For J0739+0513 (Procyon B), we rely on HST photometry
\citep{Provencal1997}.

The majority of the spectra used in this study are drawn from
SDSS\footnote{\url{https://www.sdss.org/dr14/spectro/spectro_basics/}}. The
spectra of 51 objects without SDSS spectroscopy are taken from
\citet{Limoges2013,Limoges2015}, \cite{Subasavage2007,Subasavage2009},
\cite{Bergeron1997,Bergeron2001}, or archival data secured by the
Montreal group in the last few decades. Finally, the spectroscopic
data for Procyon B are from HST \citep{Provencal2002}.

\section{Atmospheric Parameter Determination} \label{section:method}

Our fitting method is similar to that described in
\cite{Dufour2005,Dufour2007} for DQ and DZ stars,
respectively. Briefly, we first convert the observed magnitudes into
average fluxes following the method described in
\cite{Holberg2006}. The magnitudes on the AB system (SDSS, Pan-STARRS)
are converted using the relation $m = -2.5 \log{f_\nu^m} -48.60$,
where $f_\nu^m$ is the stellar flux averaged over the bandpass
response; the corrections from \citet{Eisenstein2006} are applied to
put the SDSS photometry on the AB system. The magnitudes on the Vega
system (\textit{BVRI}, {\it Gaia}) are converted using the relation $m
= -2.5 \log{f_\lambda^m} + c_m$, where $c_m$ is the zero point in the
corresponding bandpass. These zero points are calculated using the
observed fluxes and magnitudes for Vega. For the \textit{BVRI}
bandpass response, we use \cite{Cohen2003}, and for \textit{Gaia}, we
use the revised
bandpasses\footnote{\url{https://www.cosmos.esa.int/web/gaia/iow_20180316}},
while for the Vega flux, we use \texttt{alpha\_lyr\_stis\_008.fits} from the
CALSPEC Calibration
Database\footnote{\url{http://www.stsci.edu/hst/observatory/crds/calspec.html}}. Our
results are summarized in Table \ref{tab:zeropoints}.

\begin{table}
\centering
\caption{Zero points derived for different bandpasses.}
\begin{tabular}{ccc}
\hline
Bandpass & Zero point & Vega mag\\
\hline
 $G$ & $-$21.51572 & 0.00 \\
 $G_{\rm BP}$ & $-$20.99456 & 0.00 \\
 $G_{\rm RP}$ & $-$22.22677 & 0.00  \\
 $B$ & $-$20.46331 & 0.024$^1$\\
 $V$ & $-$21.06662 & 0.026$^1$\\
 $R$ & $-$21.64524 & 0.033$^1$\\
 $I$ & $-$22.37877 & 0.029$^1$\\
 \hline
(1)\cite{Holberg2006}
\end{tabular}
\label{tab:zeropoints}
\end{table}

Photometric data are also dereddened following the procedure described
in \cite{Harris2006} using the extinction maps of
\cite{Schlafly2011}. We thus consider the extinction to be negligible
for stars with $D<100$ pc, to be maximum for those located at
$|z|>250$ pc from the galactic plane ($z$ is the distance from the
galactic), and to vary linearly between these two
regimes. \cite{GenestBeaulieu2019} compared this method to that used
by \cite{Gentile2019} and came to the conclusion that they lead to
similar results for white dwarfs in the SDSS. When no parallax
measurement is available, we first obtain the photometric distance
assuming log g = 8.0, and then apply a dereddening correction using
this distance, and repeat the process until the distance (and the
reddening) converge to a single value.

Next we transform the monochromatic Eddington fluxes from our model
grids into fluxes averaged over each bandpass. These synthetic average
fluxes $H_\lambda^m$ (or $H_\nu^m$) are related to the observed fluxes
by the equation:

\begin{equation}
f_\lambda^m = 4\pi\bigg(\frac{R}{D}\bigg)^2 H_\lambda^m
\label{eq:fluxobs}
\end{equation}

\noindent
where $R$ is the radius of the star, and $D$ its distance from
Earth. In the above equation, the average model fluxes depend on
$T_\mathrm{eff}$, $\log{g}$, and the atmospheric composition, which
refers to the carbon abundance for DQ stars, and to the hydrogen and
calcium abundances for DBZ/DZ stars. The best fit between observed
photometry and synthetic fluxes is obtained using the nonlinear least
square steepest decent method of Levenberg-Marquardt
\citep{Press1986}, with the values of $T_\mathrm{eff}$ and the solid
angle $\pi (R/D)^2$ left as free parameters, while \logg and the
atmospheric composition are kept fixed during the first iteration.

From the derived value of the solid angle and the distance given by
the inverse of the parallax, we determine the radius $R$ of the
star. The values of $\log{g}$ and mass are then derived by
interpolating in evolutionary models similar to those described in
\cite{Fontaine2001} but with C/O cores, $\log{q(\textrm{He})}\equiv
\log M_{\rm He}/M_{\star}=-2$ (where $M_\star$ is the mass of the star) and
$\log{q(\textrm{H})}=-10$, which are representative of helium-rich
atmosphere white dwarfs \citep{Dufour2005}. We repeat the process this
time using the newly determined surface gravity until convergence is
reached. If the parallax is unknown, we simply assume $\log{g}=8$ and
derive a radius $R$ from the same evolutionary models, which yields a
photometric distance when combined with the solid angle. The
uncertainties on the effective temperature and the solid angle are
obtained directly from the covariance matrix of the fitting procedure.

We next determine the chemical composition by fitting the
spectroscopic data, again using the Levenberg-Marquardt method,
keeping \teff and \logg fixed to the values obtained from the
photometric fit. For cool DQ stars, we fit the Swan bands between 4000
and 6500 {\AA}, while for hotter DQs, we use the atomic absorption
lines between 4500 and 5500 {\AA}.  The carbon abundance, the solid
angle, and a first or second degree polynomial --- to account for
uncertainties in the flux calibration \citep[see][]{Dufour2005} ---
are considered free parameters during the fitting procedure. For
DBZ/DZ stars, we begin by fixing the hydrogen abundance, either by
fitting the H$\alpha$ spectral line if it is visible, or by fixing
\loghhe at the detection limit (we also fit stars not showing
H$\alpha$ with our hydrogen-free grid; see Section \ref{Section:DZ}
for a discussion of this matter). This detection limit was estimated
by calculating, at each temperature and $\log{g}$ value in our grid,
the amount of hydrogen required to reach a threshold of 500 m{\AA} for
the equivalent width of the H$\alpha$ line. We then fit the Ca
\textsc{ii} H\&K absorption lines to determine \logcahe. Then we
repeat the photometric fit, but this time using the values of
$\log{g}$, calcium abundance (or carbon abundance in the case of DQ
white dwarfs), and hydrogen abundance obtained in this last
iteration. We repeat the procedure, typically 3 to 5 iterations (or
more in some cases), until the parameters have converged to a stable
solution.

To obtain the uncertainty on the abundances measured from the
spectroscopic fit, we rely on the same method described in
\cite{Bergeron1992}, where we first assume an arbitrary standard
deviation $\sigma = 1$ for each data point, then calculate the
root-mean-square deviation of the observed spectrum from the best-fit
model spectrum. This is then propagated into the covariance matrix,
from which the formal uncertainties of the fitted atmospheric
parameters are obtained. This gives an uncertainty estimation that
depends mostly on the signal-to-noise ratio of the observed spectra.

\section{DZ/DBZ(A) White Dwarfs} \label{Section:DZ}

\subsection{Photometric and Spectroscopic Fits} \label{section:dzfits}

\begin{figure*}
	\centering
    \includegraphics[width=1.5\columnwidth]{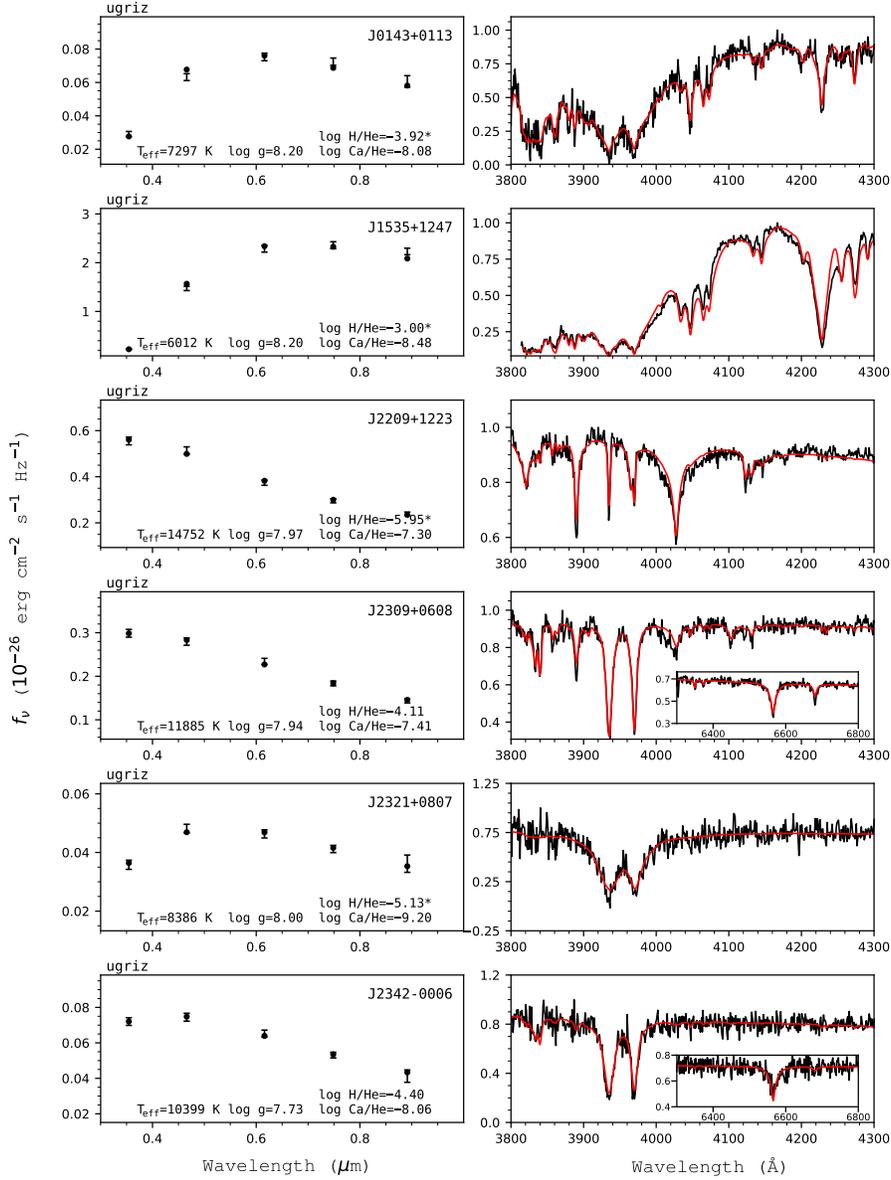}
    \caption{Examples of fits to our sample of DBZ/DZ(A) white
      dwarfs. Left panels: Photometric fits where error bars represent
      the observed data, while filled circles correspond to average
      model fluxes. A dagger symbol indicates that \logg is fixed at
      8.0 (no parallax measurement available), while a star symbol
      indicates a value of \loghhe fixed at the visibility
      limit. Right panels: Spectroscopic fits (red) to the normalized
      observed spectra (black). The inset shows the fit to H$\alpha$
      when present. A version of this figure with all our fits is
      available in Appendix I.}
    \label{fig:dzexample}
\end{figure*}

Following the method described in Section \ref{section:method}, we
determined the atmospheric parameters for all DZ/DBZ(A) white dwarfs
in our sample. Figure \ref{fig:dzexample} shows typical fits of the
energy distribution and the Ca \textsc{ii} H\&K lines region (all our
fits are available in Appendix I), while our final parameters are
given in Table \ref{tab:resultsdz}. The fit to the \halpha line, when
present, is shown as an inset. Note that for stars that do not show
hydrogen, we report a solution with the hydrogen abundance fixed at
the detection limit (see explanation below).

We note that other lines, mostly Mg and Fe, are sometimes observed in
addition to Ca \textsc{ii} H\&K. Since in our model grids, the
abundances of these elements relative to calcium are fixed to the
chondrite values, a visual inspection of the quality of the ``fit'' to
these lines provides a quick assessment of the validity of this
approximation, which appears to be adequate (or close enough) for most
stars that show such lines in our sample. However, it is certainly
possible that some elements depart significantly from the chondrite
values. Since what is most important, as far as the mass and effective
temperature determinations are concerned, is the amount of free
electrons in the photosphere (effects due to the redistribution of the
flux absorbed in the UV can also play a role in some cases), small
deviations should only have a minimal impact on those parameters, even
more so when hydrogen is present, because electrons from ionized
hydrogen will dominate the free electron budget.

Recently, \cite{Hollands2017,Hollands2018} analyzed a large sample of
230 cool DZ white dwarfs and determined abundances of individual
elements. Their results indicate that to first order, objects are
found with chemical compositions not too far from that of bulk Earth
ratios, at least for the most visible elements, which are calcium,
magnesium, and iron. Although their parameters were obtained using a
different model atmosphere code that do not include all the
improvements included in our code, a comparison of their results for
197 objects in common with our sample suggests that the exact
metal-to-metal ratios used in their analysis has only a modest impact
on the effective temperature determination. We find that the effective
temperatures obtained with our approach are, on average, only 11 K
lower than theirs, but with a somewhat large 330 K standard
deviation. We find no correlation between the differences in effective
temperature and the deviation to the chondrite abundances (relative to
Ca) that they report. Instead, we find that these differences are
correlated with $\log g$ values. Since the {\it Gaia} second data
release was not available at the time the analyses of
\cite{Hollands2017,Hollands2018} were published, the authors assumed
\logg $=8$ for their whole sample, while our analysis makes use of the
newly available parallax measurements to determine the surface
gravities. This distinction seems to be the main explanation for the
different atmospheric parameters derived in both studies. Indeed, we
find a strong correlation between the differences in effective
temperature and the departure from \logg $=8$, our temperatures being
higher (lower) for larger (smaller) $\log g$ values. Similarly, there
is also a correlation between the differences in log Ca/He and $\log
g$. This result is not surprising given that the difference in the
strength of the calcium absorption features resulting from a change in
surface gravity will need to be compensated by an appropriate change
in calcium abundance, which in turn will affect the number of free
electrons, and thus the effective temperature determination. We thus
conclude that the atmospheric parameters derived using our approach
should be reliable, and not significantly affected by our assumption
of the metal-to-metal ratios.

Nevertheless, when better spectroscopic data (high-resolution or UV
observations) for a given object indicate some departure from our
approximation, atmospheric parameters should always be re-derived in a
self-consistent way for better precision. For example, for the DZA
white dwarf Ross 640 analyzed by \citet{Blouin2018a}, changes in the
abundances of Mg and Fe relative to Ca affected the UV flux level
sufficiently to warrant the calculations of a specific grid with
modified abundances in order to obtain atmospheric parameters in a
self-consistent way. As a result, their final effective temperature
was 250 K cooler, and the corresponding mass 0.04 $M_{\odot}$ lower,
than what we obtain here with our approach. Unfortunately, since
high-resolution or ultraviolet spectroscopic data are not likely to
become available for every object in our sample in a foreseeable
future, we are forced to adopt this approximation, in particular given
the fact that most objects in our sample show only calcium lines. To
conclude on this topic, while our approach provides the best
atmospheric parameters that are possible to infer with the available
data, the solutions for individual objects will always suffer from
small intrinsic uncertainties related to our adopted metal abundances.

\subsection{Hydrogen Abundance Measurements} \label{section:hydrogen}

While for most objects in our sample, the metal-to-metal ratio assumed
in our analysis has only a modest impact on our atmospheric
parameters, the abundance of hydrogen, on the other hand, has a deeper
impact because it can be one of the main free electron donors, even
when present below the visibility limit. For instance,
\citet{Bergeron2019} recently showed that adding undetectable traces
of hydrogen in the models had a non-negligible effect on the mass
determination of helium-rich white dwarfs (this is also discussed in
\citealt{Dufour2005} in a similar context, but with carbon as the main
electron donor).

Only 105 of the 1023 (10\%) white dwarfs in our sample show H$\alpha$
in their spectrum. This is much less than the 25\% containing hydrogen
reported by \cite{Dufour2007} for two main reasons. The first one is
that the presence of hydrogen for 7\% of their objects was determined
indirectly from the shape of the Ca \textsc{ii} H\&K absorption
features, in the sense that much better fits to these lines could be
achieved when hydrogen was included. However, the study of Dufour et
al.~was based on Lorentzian profiles, with \logg fixed at 8.0, while
here we use the unified line shape theory of \citet{Allard1999}, with
surface gravities constrained by parallax measurements. As a
consequence, we no longer find objects with spectroscopic fits that
are significantly improved by adding hydrogen, indicating that the
need to add hydrogen in the Dufour et al.~analysis was probably only a
way to compensate sub-optimal line profiles and/or incorrect surface
gravities. The second reason is that since Dufour et al., the
proportion of DZ white dwarfs too cool to show hydrogen has increased
substantially, thanks to the thorough search for metal-polluted white
dwarfs near the main-sequence color space \citep{Koester2011,
  Hollands2017, Hollands2018}.

\begin{figure}
\centering
   \includegraphics[width=\columnwidth]{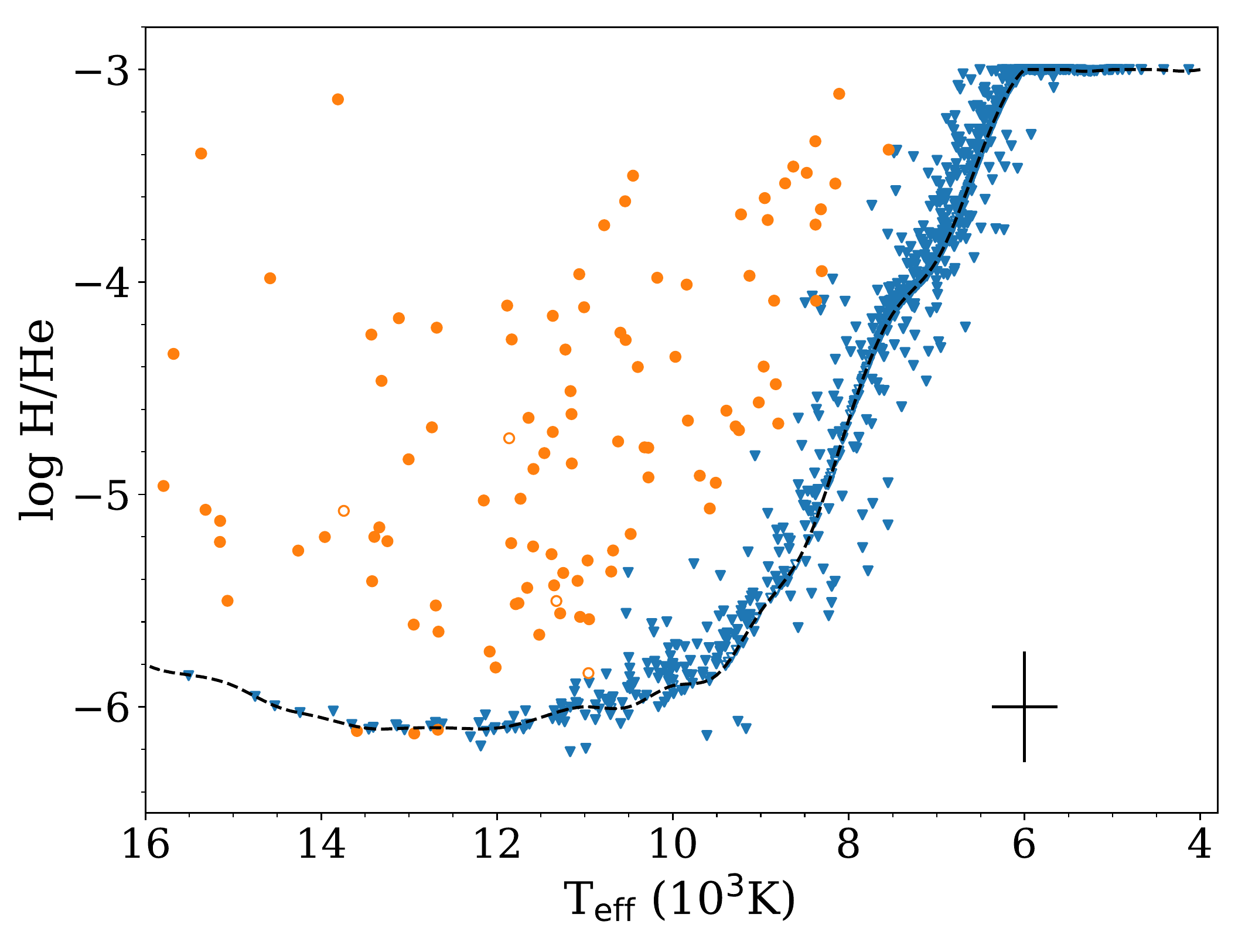}
    \caption{Hydrogen abundance as a function of effective temperature
      for our sample of DBZ/DZ(A) white dwarfs. Objects with parallax
      measurements are represented with filled symbols, while open
      symbols are used for stars for which we assumed log g =
      8. Orange circles are objects with hydrogen abundances
      determined from fitting H$\alpha$, while blue triangles
      correspond to upper limits. The dashed line represents the
      visibility limit, defined as an equivalent width of 0.5 {\AA}
      for \halpha at $\log{g}=8$. The black error bars represent the
      average uncertainties.}
  \label{fig:loghhe}
\end{figure}

\begin{figure}
	\centering
    \includegraphics[width=\columnwidth]{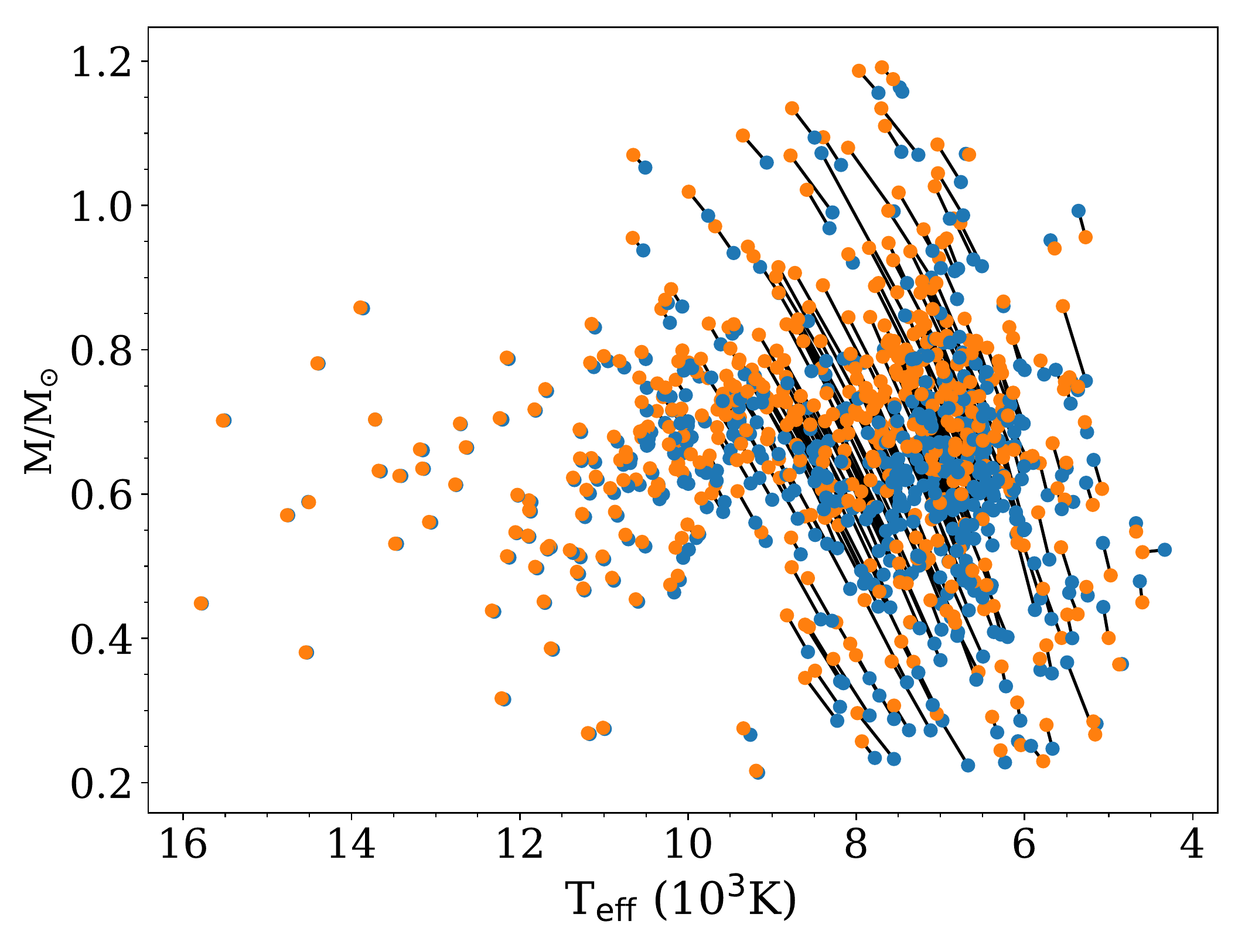}
    \caption{Comparison of masses and effective temperatures obtained
      with hydrogen-free models (orange dots) and with models where
      the hydrogen abundance was set at the visibility limit (blue
      dots). A black line connects the two solutions for each object.}
  \label{fig:photocomp_indiv}
\end{figure}

Figure \ref{fig:loghhe} shows the abundance of hydrogen as a function
of effective temperature for all the DBZ/DZ(A) white dwarfs in our
sample. For stars that do not show H$\alpha$, we determine the maximum
amount of hydrogen that can be added without being detected (the
scatter is explained by variations in \logg for each object).  While
the hydrogen content, and thus its impact on the free electron budget,
is well constrained for $T_{\rm eff}\gtrsim9000$ K, increasingly large
quantities of hydrogen can be hidden as the effective temperature
decreases. To examine the impact of the unknown amount of hydrogen on
our atmospheric parameter determinations, we fit each star in our
sample with no detectable \halpha with both a hydrogen-free model
grid, and with the abundance of hydrogen fixed at the detection
limit. The results of this experiment are displayed in Figure
\ref{fig:photocomp_indiv}.

\begin{figure}
	\centering
    \includegraphics[width=\columnwidth]{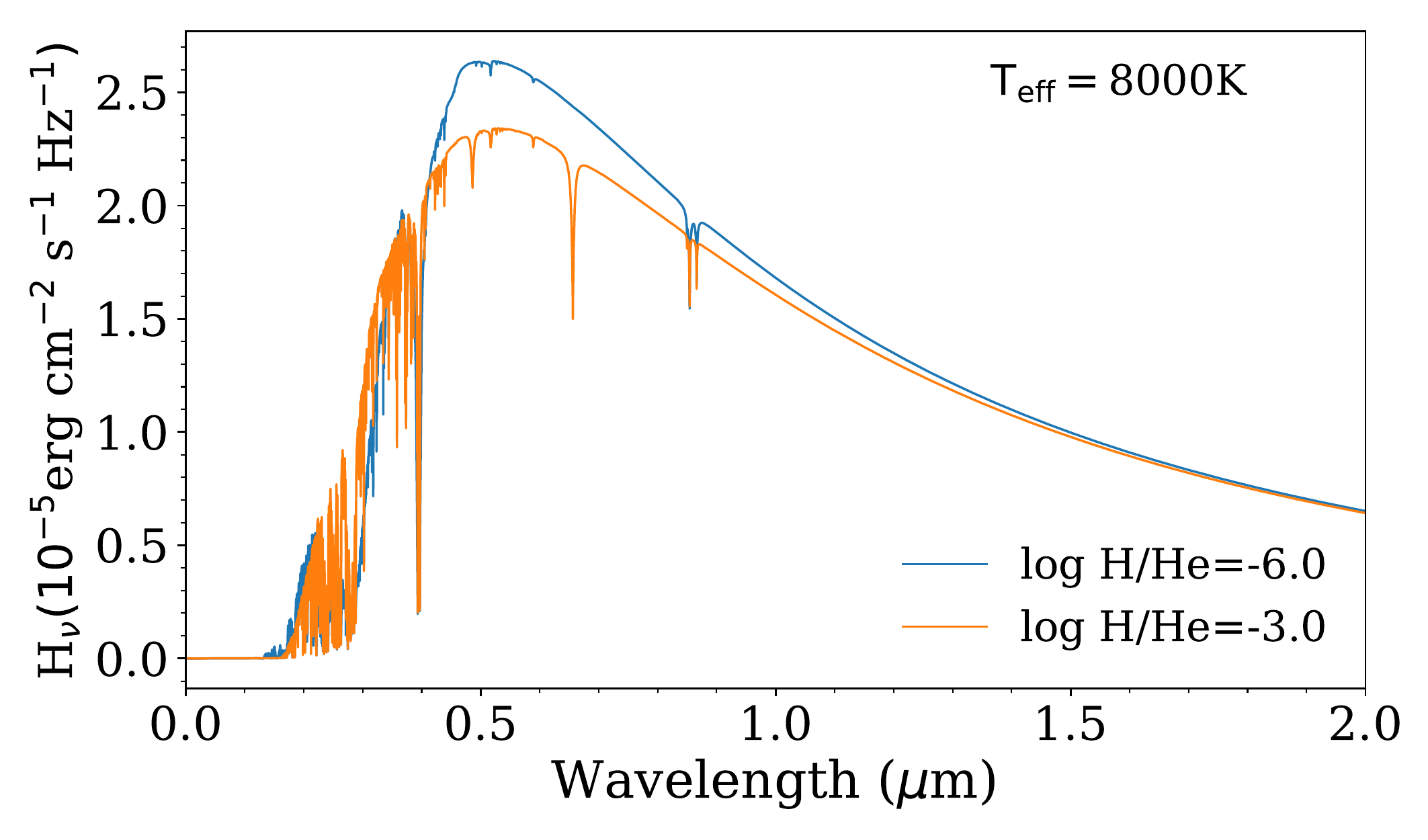}
    \caption{Synthetic spectra (Eddington fluxes) for models at
      $T_{\textrm{eff}} = 8000$ K, $\log{\textrm{Ca}/\textrm{He}}=-9$,
      and $\log{g}=8$, with two different hydrogen abundances
      indicated in the figure.}
  \label{fig:comparisonhhe}
\end{figure}

Clearly, masses and effective temperatures are significantly reduced
for the coolest stars when hydrogen is included. This reduction in
both effective temperature and mass was recently explained by
\citet[][see their Figures 10 and 11]{Bergeron2019}. Briefly, adding
hydrogen in the model increases the number of free electrons, which in
turn increases the He$^{-}$ free-free opacity. This has a quite
dramatic effect on the continuum, as can be appreciated from Figure
\ref{fig:comparisonhhe}, where the energy distribution of two models
that differ only by their hydrogen content are shown. As a result of
the increased He$^{-}$ free-free opacity, a lower temperature and a
larger solid angle are required to match the observed fluxes, which
translate into a larger radius, and thus a smaller mass. Note that
this effect practically disappears for effective temperatures above
$\sim$11,000 K, as the contribution from ionized helium starts to
dominate the free electron budget.

\subsection{Mass Distributions} \label{section:dzmdistr}

As discussed above, the parameters we derive depend intimately on the
assumed amount of invisible hydrogen present in the star. It is
certainly reasonable to expect at least some amount of hydrogen to be
present in each object. After all, not only have these white dwarfs
been traveling through the interstellar medium for billions of years,
but potentially, they also have accreted bodies that may have
contained large amounts of water and ice \citep{Klein2010, Farihi2011,
  Farihi2013,Raddi2015}. While it is impossible to determine exactly
the hydrogen abundance in each object, we know this abundance must lie
within the two limiting cases explored here. We thus, in what follows,
take a deeper look at the consequences this parameter has on the
derived global properties of the sample, and most importantly on the
mass distribution.

\begin{figure}
	\centering
	\includegraphics[width=\columnwidth]{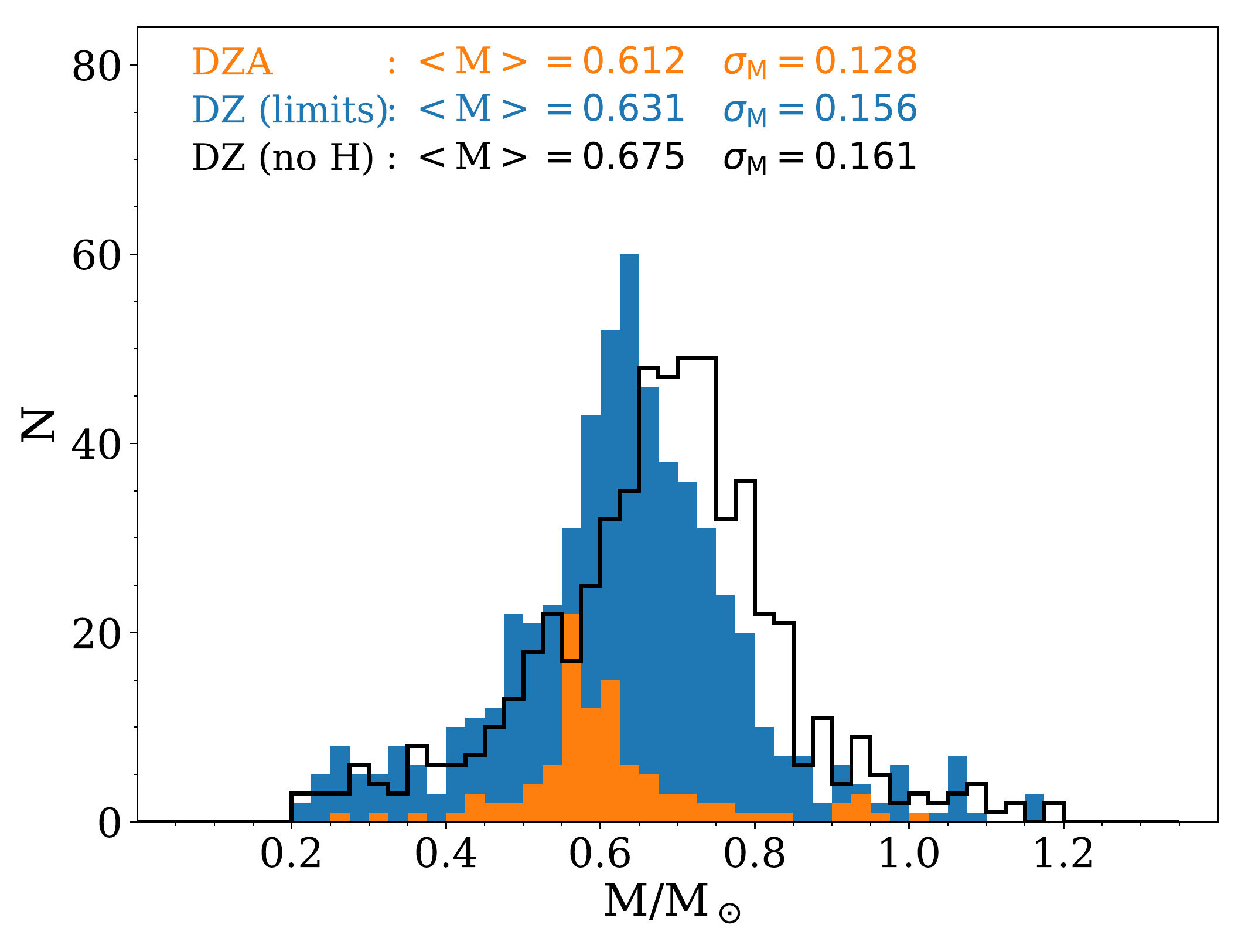}
  \caption{Mass distributions for DZ and DZA white dwarfs in our
    parallax sample. The DZA stars are shown in orange, while the DZ
    stars fitted with hydrogen-free models or with the hydrogen
    abundance set to the visibility limit are shown in black and blue,
    respectively.}
  \label{fig:histomass}
\end{figure}

Figure \ref{fig:histomass} compares the mass distributions of our
parallax sample for stars with hydrogen abundances measured directly
from H$\alpha$ (105 objects) with those with no detectable H$\alpha$
feature (918 objects) analyzed both with hydrogen-free models and with
hydrogen abundances set to the visibility limit. We find that the mean
mass for the DBZA/DZA stars is 0.612 \msun, very close to the value
recently reported by \citet{GenestBeaulieu2019} for DA (0.617 \msun)
and DB (0.620 \msun) white dwarfs. However, the mean mass is
significantly higher (0.675 \msun) for stars without H$\alpha$ if no
hydrogen is included in our models. This indicates that the effective
temperatures and masses obtained from hydrogen-free models are
probably overestimated, most likely due to the presence of invisible
traces of hydrogen in these stars. Adding the maximum amount of
hydrogen pushes the mass distribution towards values that now become
consistent (mean mass of 0.631 \msun) with the mean mass of DA and DB
white dwarfs, although the peak is still slightly shifted towards
higher masses with respect to the DBZA/DZA stars.

Since there is no reason to believe that these objects represent a
distinct massive population, this probably means that our effective
temperatures and masses are still slightly overestimated for many
objects using our approach, most likely due to small deviations of the
abundances of some elements with respect to the chondrite values. For
example, if the accretion episode responsible for the metal pollution
in some of these stars has stopped a long time ago, elements with
different masses will start to settle with different timescales
(photospheric abundances will be very close to that of the polluting
body only during the early phase and steady state phase;
\citealt{Dupuis1993, Koester2009}), leading to abundances that can
depart significantly from our assumed ratios. This will also affect,
to a lesser extent, the relative strength of the absorption in the
UV. Consequently, the uncertainties on the atmospheric parameters for
these stars will be intrinsically larger than the statistical values
reported here. Spectroscopic observations in the ultraviolet for every
object in our sample would be required for a more accurate analysis,
but unfortunately, such data will not become available anytime
soon. Note that DZA white dwarfs are less affected by these
uncertainties because when hydrogen is present in sufficient
quantities to be detectable, it will also be the main free electron
donor at the photosphere, diminishing the impact of small variations
of heavy element abundance ratios with respect to chondrites.

Even though it is not possible to derive the exact photospheric
hydrogen abundance for each object, the fact that the peak of the mass
distribution appears more realistic when traces of hydrogen are
included suggests that most of these objects probably have abundances
close to the detection limit. As will be discussed in Section
\ref{Section:spectralevolution}, this corresponds also approximately
to the amount of hydrogen expected if such DZ stars are the results of
convectively mixed DA white dwarfs. Consequently, our adopted
atmospheric parameters reported in Table \ref{tab:resultsdz} are
obtained with the hydrogen abundance set to the visibility limit,
keeping in mind that the true solution for individual objects may be
off by a few hundred degrees and a few hundredth solar mass, depending
on their real hydrogen (and also metal) content.

Note finally that objects in Table \ref{tab:resultsdz} with $M\lesssim
0.47$ \msun\ are most likely unresolved double degenerate binaries,
because such low mass white dwarfs would have low mass progenitors on
the main sequence, with lifetimes longer than the age of the Milky
Way. Such binaries are more luminous, resulting in a larger solid
angle for a given distance, and thus a larger radius (lower mass) is
inferred when analyzed under the assumption of a single object.

\subsection{Accreted Material}

\begin{figure}[t]
\centering
    \includegraphics[width=\columnwidth]{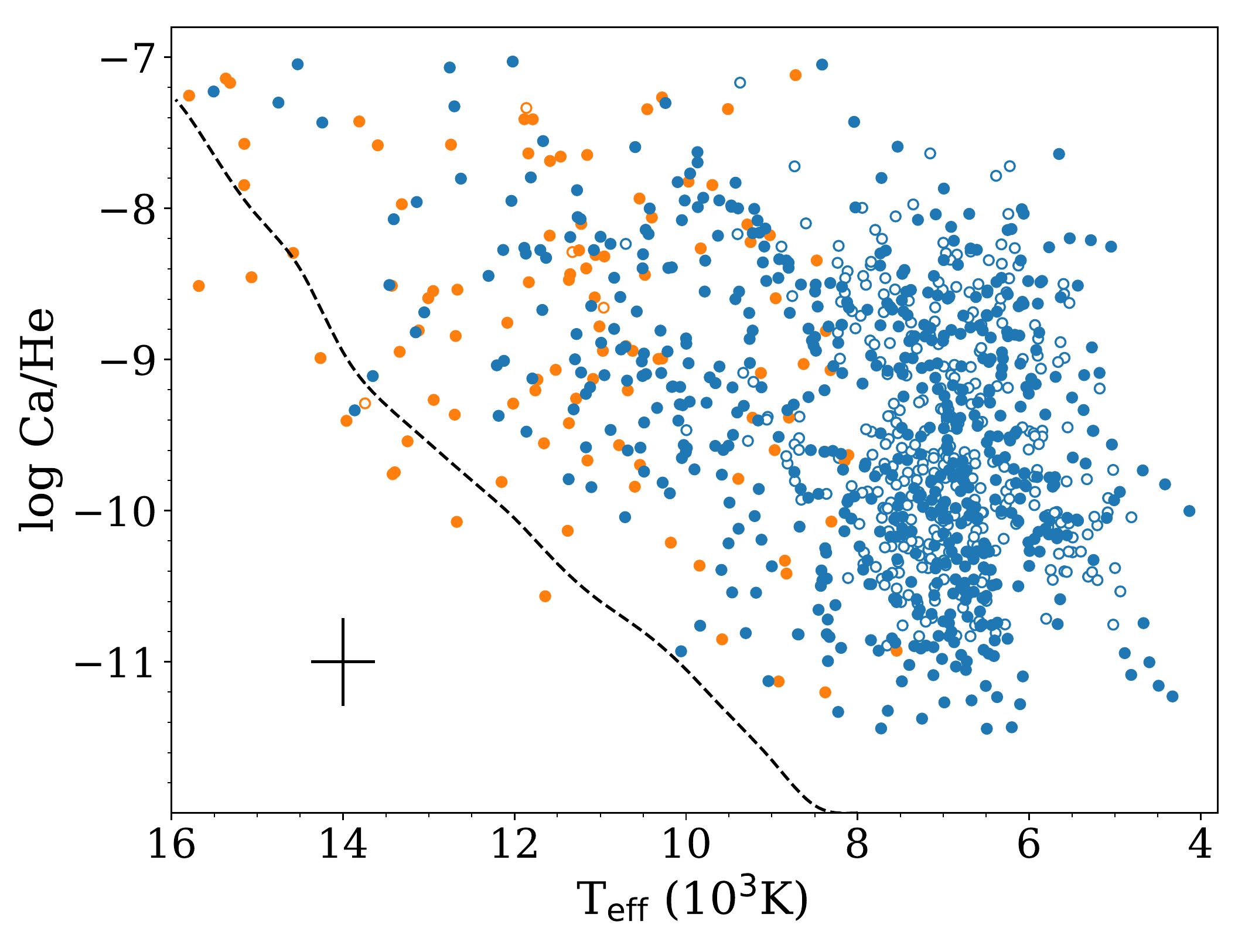}
    \caption{Calcium abundance as a function of effective temperature
      for all DZ white dwarfs in our sample. The black cross indicates
      the mean error bar. White dwarfs with a parallax measurement are
      shown with filled symbols, while $\log g = 8$ was assumed for
      objects with open symbols. Orange circles represent DZA stars
      with hydrogen abundances determined by fitting H$\alpha$, and
      blue circles represent DZ stars with the hydrogen abundance
      fixed at the detection limit. The dashed line indicates the
      detection limit of calcium, defined as an equivalent width of
      0.5 {\AA} for the Ca \textsc{ii} H line.}
  \label{fig:logcahe}
\end{figure}

Figure \ref{fig:logcahe} shows the abundance of calcium as a function
of effective temperature for the 1023 DBZ/DZ(A) white dwarfs in our
sample. The two gaps in the distribution reported by
\cite{Dufour2007}, namely between 5000 and 6000 K as well as in the
top right corner of the diagram, have been mostly filled, thanks to
the recent discovery of many new cool polluted white dwarfs
\citep[][]{Kepler2015,Kepler2016,Kleinman2013, Koester2011,
  Hollands2017, Hollands2018}. The absence of objects at high
effective temperatures and low calcium abundances is still present,
and is due to the detection limit of Ca \textsc{ii} H\&K lines, as
indicated in the figure. At the typical resolution of our
spectroscopic observations, stars with such parameters would simply
appear as DC white dwarfs below \teff $\sim$ 12,000 K, and as DB stars
above this temperature.

\begin{figure}
	\centering
    \includegraphics[width=\columnwidth]{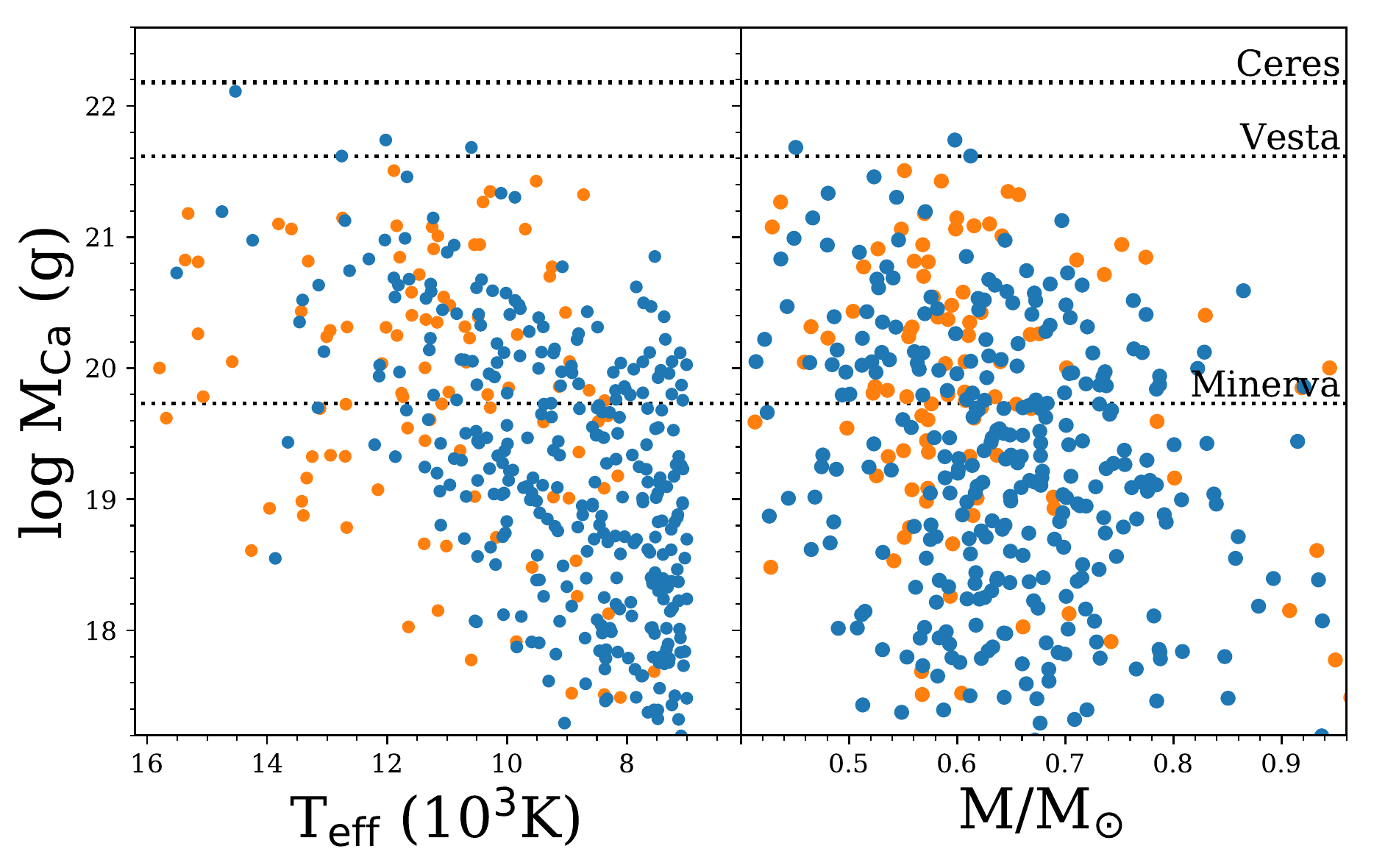}
    \caption{Total mass of calcium in the convection zone as a
      function of effective temperature (left) and mass
      (right). Orange circles are objects with hydrogen abundances
      determined by fitting H$\alpha$ and blue circles are objects
      with the hydrogen abundance fixed at the detection limit. We
      also show the estimated mass of calcium in Ceres, Vesta, and
      Minerva, assuming that calcium constitutes 1.6\% of their total
      mass as in bulk Earth \citep{Farihi2011}.}
  \label{fig:calciummass}
\end{figure}

More interesting than the calcium-to-helium abundance ratio is the
total mass of calcium contained in the convection zone of these white
dwarfs. Using the derived effective temperature and surface gravity of
each object, we can calculate the total mass of the helium convection
zone using envelope models similar to those described in \citet[][see
  \citealt{Dufour2010} for further details]{Fontaine2001} ranging from
$T_\mathrm{eff} = 7000$ K to $30,000$ K, $\log{g}=7.5$ to $9.0$, and
with the ML2/$\alpha=1.0$ version of the mixing-length theory (the
effect of changing the mixing length is negligible below $T_{\rm
  eff}\sim16,000$ K; see Figures 9 and 10 of
\citealt{Rolland2018}). Our results are presented in Figure
\ref{fig:calciummass}. Note that these values represent only lower
limits to the total mass accreted since we do not know how much mass
has already diffused at the bottom of the convection zone. Also shown
for comparison in Figure \ref{fig:calciummass} is the estimated mass
of calcium present in well-known asteroids in the solar system,
assuming that calcium accounts for 1.6\% of the mass of a typical
planetesimal \citep[see][]{Farihi2011}. Our results are similar to
those presented in \cite{Farihi2011}, with masses ranging from
$10^{18}$ g to $10^{22}$ g, with possibly smaller mass values at lower
effective temperatures (the lack of low calcium mass values at high
effective temperatures is simply a visibility limit effect; see Figure
\ref{fig:logcahe}). Surprisingly, the total mass of calcium remains
constant as a function of the white dwarf mass between 0.5 \msun\ and
0.75 \msun, with a possible decrease in more massive white dwarfs,
although we are probably dealing here with small number statistics.

\subsection{Main Sequence Progenitors} \label{section:msprogen}

\begin{figure}
	\centering
    \includegraphics[width=\columnwidth]{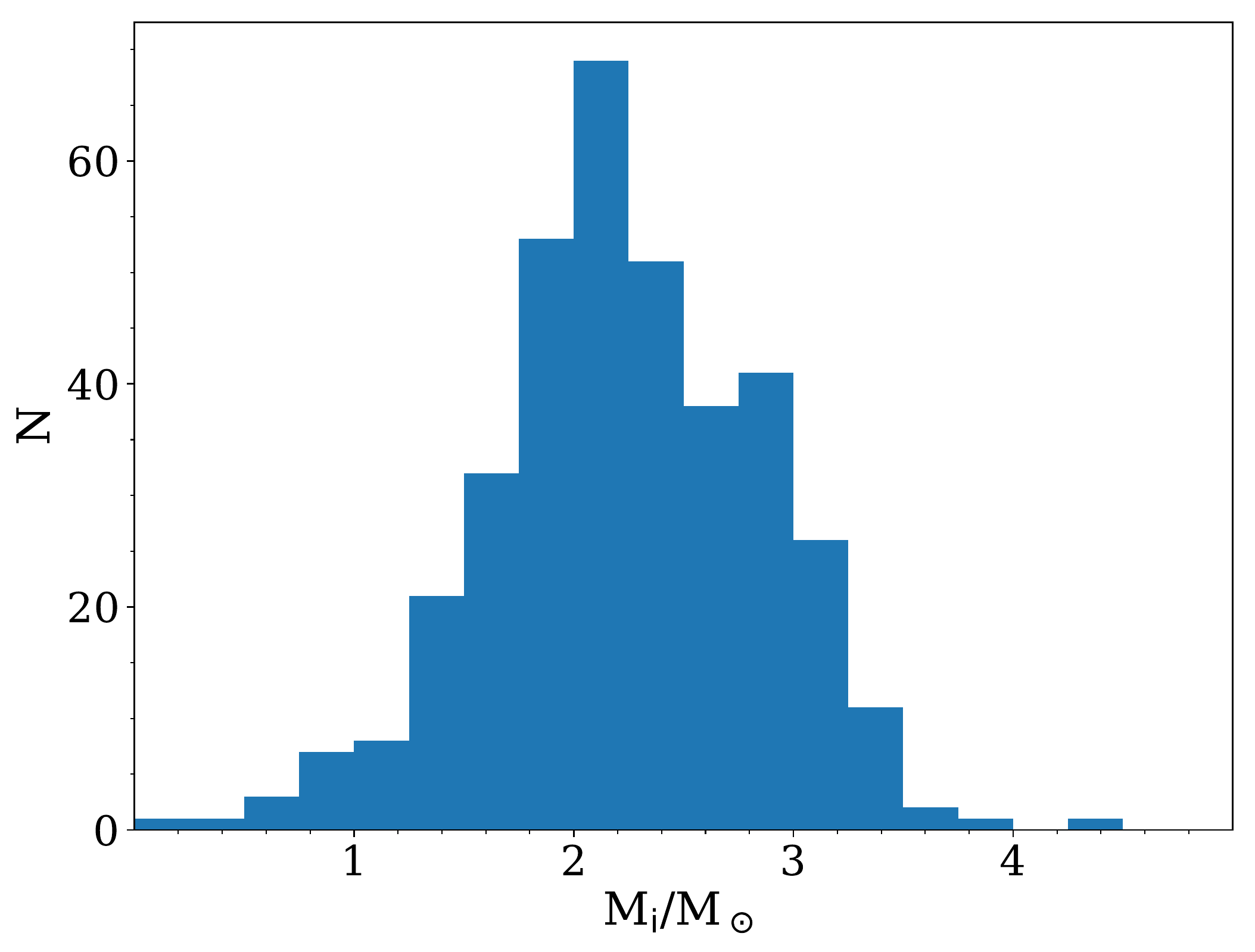}
    \caption{Mass distribution of white dwarf progenitors based on the
      IMFR of \cite{Williams2009} for the DBZ/DZ(A) stars in our
      sample with $\sigma_\pi/\pi<0.1$.}
  \label{fig:initialmass}
\end{figure}

It is also possible to estimate the mass of the white dwarf
progenitors using the empirical initial-final mass relation
(IFMR). While there are many different IFMR published (see Figure 8 of
\citealt{Williams2009} for a comparison), most of them rely on the
observations of star clusters. \cite{El-Badry2018} propose an
alternative way of calculating the relation using the {\it Gaia}
color-magnitude diagram. Their results agree remarkably well with
those of \cite{Williams2009}, which we decided to use for
simplicity. One must keep in mind that this relation includes many
systematic errors, one being the possible dependence on initial
metallicity \citep{Marigo2007}, information that is lost when
evolving to the white dwarf phase. Because of this, it is difficult to
determine accurate initial masses of individual objects, but we can
obtain a good idea of the distribution. The result is shown in Figure
\ref{fig:initialmass} for stars with $\sigma_\pi/\pi<0.1$.  Note that
below $M_i = 1$ \msun, the results are meaningless, since the
white dwarfs in these bins have masses below 0.47 $M_\odot$, and are
most probably unresolved double degenerate binaries (single white
dwarfs with such low mass have long main sequence lifetime, and are
not expected to have evolved into white dwarfs within the age of the
galactic disk yet).

On the basis of the results displayed in Figure \ref{fig:initialmass},
we find that 41 white dwarfs had main sequence progenitors with masses
above 3 \msun, indicating that the formation of rocky objects around
massive stars may not be exceptional at all. Note that the {\it
  Extrasolar Planets
  Encyclopaedia}\footnote{\url{http://exoplanet.eu/}} reports only 9
exoplanets (out of 3802 exoplanets with the mass of the host star
determined) around 8 different stars with masses above 3 \msun. The
very small number of known planets around massive stars is mostly due
to various selection effects, as the method used to find them
(transit, radial velocity, direct imaging) are all less efficient
around large, bright, and massive main sequence stars. Nevertheless,
theoretical models of planet formation predict that planet occurrence
around massive stars should be higher, and the presence of rocky
material in the photosphere of many polluted white dwarfs with massive
main sequence progenitors seems to confirm this hypothesis. However,
to properly study planet occurrence and its correlation with stellar
mass, an analysis of a large sample of DC and DB white dwarfs would be
necessary, preferably for a complete-volume sample, something that is
outside the scope of this work.

\subsection{Discussion of Individual Objects} \label{section:dzobjects}

\textit{J0005+7313} --- The helium lines at 3889 and 4026 {\AA} are
deeper than those predicted by the model, indicating that the
temperature determined from photometry, \teff $=$ 12,673 K, is
probably underestimated. Using {\it Gaia} photometry, we find \teff
$=$ 13,152 K, which is still too cool for a good spectroscopic
fit. \cite{Bergeron2011} found \teff $=$ 14,410 K, \loghhe $=-5.97$,
and no metals, using only spectroscopy. The absence of metals is not
the explanation, since we obtain 12,715 K fitting photometry with our
metal-free models. Because their technique is independent of
photometry, one possible explanation is the reddening due to
interstellar absorption. In this work, we use the procedure of
\cite{Harris2006}, and since the star is only at 34.7 parsecs, no
correction for reddening is applied. If we use the procedure of
\cite{Gentile2019} instead, we need to apply 3.2\% of the maximum
absorption along the line of sight, and we now obtain \teff $=$ 13,622
K, which provides a much better agreement with the spectra. Three
dimensional reddening maps made with {\it Gaia} should eventually
allow better correction for reddening.

\textit{J0152+2418} --- The Mg \textsc{i} lines at $\sim$3830 {\AA}
and the red wing of the Ca \textsc{ii} H\&K lines are not well
reproduced by our model. One possible explanation is that the
abundance ratios differ significantly from that of chondrites. A
deficiency in magnesium (and possibly Fe and other elements) may
explain this discrepancy. More specific adjustments of the various
abundances would be necessary to obtain a good fit.

\textit{J0209+2914}, \textit{J1242+0829}, and \textit{J1424+5657} ---
All three white dwarfs have ``flattened'' Ca \textsc{ii} lines,
similar to those observed in J1249+6514, not analyzed here, which has
been identified as magnetic by \cite{Hollands2017}. The spectra do not
show other lines that could be used to detect line splitting. We can
only consider the values we found to be approximate, but if the
presence of magnetism is confirmed, J1242+0829 would be the hottest
known magnetic DZ star with \teff $\sim$ 8123 K.

\textit{J0302$-$0108 (GD 40)} --- Our effective temperature of \teff
$=$ 13,594 K is much lower than that reported by \cite{Voss2007} ---
\teff $=$ 15,316 K (with \logg fixed at 8) --- obtained on the basis
of metal-free models. They also relied on optical spectra rather than
photometry to derive the temperature. Part of this large difference is
probably attributable to the lack of heavy elements in their model
atmosphere calculations, but since our predicted helium lines are a
tiny bit too shallow compared to the observations, it is also possible
that our effective temperature is slightly underestimated, possibly
due to our neglect of any reddening correction given that the distance
is only 64 pc.

\textit{J0555$-$0410 (LP 658-2)} and \textit{J2201+0219} --- While
these objects do not show a clear H$\alpha$ absorption line, setting
the hydrogen abundance at the detection limit leads to completely
spurious fits, so we decided to adopt hydrogen-free models for these
stars. J0555$-$0410 was also analyzed in \cite{Blouin2018a} who also
found a lower hydrogen abundance of \loghhe $<-5$. The difference of
61 K in \teff between both temperature estimates can be explained by
the different sets of photometry used (\textit{BVRI}+\textit{JHK}
instead of Pan-STARRS \textit{grizy}). Blouin et al.~were also able to
constrain $\log{\mathrm{Mg}/\mathrm{He}}$ to $-8.66$, a value much
higher than the $-9.92$ we used here on the basis of the chondrite
ratio.

\textit{J0801+5329}, \textit{J0842$-$1347}, and \textit{J1428+4403}
--- These are objects similar to \textit{J0152+2418} where the
magnesium lines and the shape of the Ca \textsc{ii} H\&K lines are not
well reproduced. Reducing \loghhe improves the fit in one case
(J1428+4403), but not to the point where the fit is
satisfactory. Additional adjustments of individual metal abundances
are most probably needed for these objects.

\textit{J0846+3538} and \textit{\textit{J1356+4047}} --- Both stars
seem to have an overabundance of magnesium and sharper than predicted
Ca \textsc{ii} H\&K lines, as opposed to J0152+2418 and other similar
objects discussed above. Again, a fit using a tailor-made grid with
different abundance ratios would probably allow a better fit for these
objects.

\textit{J1214+7822} --- It was impossible to obtain a good
spectroscopic fit for both the Ca \textsc{ii} H\&K lines and the Ca
\textsc{i} line at 4226 \AA\ simultaneously. We found the best
agreement with H\&K by using hydrogen-free models and ignoring the Ca
\textsc{i} line. However, with a mass of only 0.31 \msun, this star is
most likely a double degenerate system, and a DZ+DC system could
explain the unusually narrow H\&K spectral lines at this
temperature. Alternatively, \cite{Limoges2015} suggested that this
object could have a hydrogen-rich atmosphere, with a lower atmospheric
pressure and thus narrower absorption lines.

\textit{J1234+5606} --- The SDSS magnitudes for this
object are $\sim$0.3 mag fainter than Pan-STARRS and {\it Gaia}, which
leads to two different possible solutions. We could not reach a
conclusion regarding this discrepancy, but the SDSS colors lead to a
much better spectroscopic fit of the helium lines at 5876 and 6678
\AA, indicating a good estimate of the effective temperature. We thus
decided to adopt the SDSS photometric data set.

\textit{J2253$-$0646 (WD 2251$-$070)} --- \cite{Blouin2019b} found for
this object \teff $=4170 \pm 90$ K, \logg $=8.06 \pm 0.08$, and
\logcahe $=-9.8 \pm 0.2$ from fitting \textit{BVRI}+\textit{JHK} and
Pan-STARRS photometry, while we find \teff $=4132 \pm 53$ K, \logg
$=8.03 \pm 0.07$, and \logcahe $=-10.00 \pm 0.05$. The values are very
close, but \cite{Blouin2019b} show a much better fit in their Figure
9. This is not surprising because they relied on improved line profile
calculations for the Ca \textsc{i} line at 4226 {\AA}. These improved
calculations are important only for objects with \teff $\lesssim 4500$
K, and thus they have little impact on our analysis. Only 4 objects in
our sample are in that temperature range; J1636+1619 and J0555$-$0410
do not show Ca \textsc{i} line, and J1214+7822 has already been
discussed above.

\section{DQ White Dwarfs} \label{Section:DQ}

\subsection{Carbon Abundances} \label{Section:Cdistr}

\begin{figure*}
	\centering
    \includegraphics[width=1.5\columnwidth]{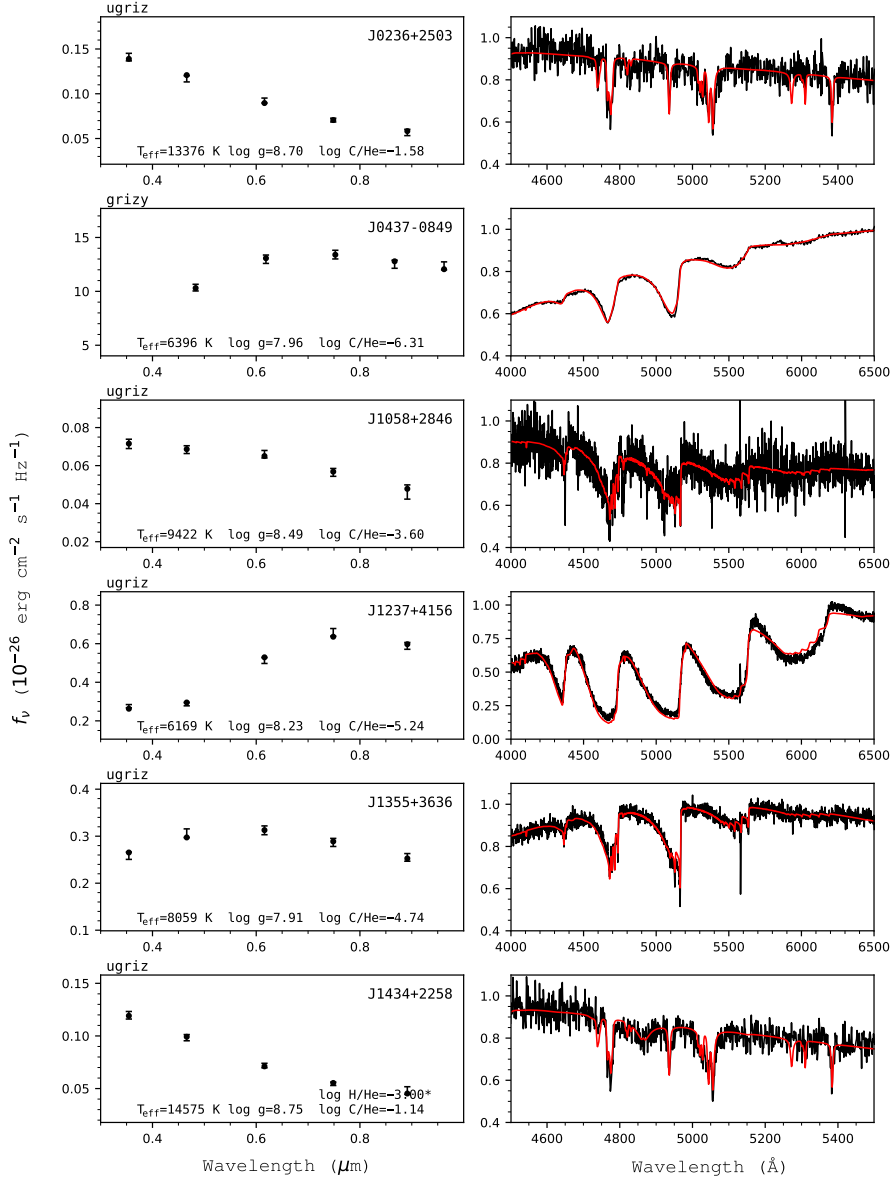}
    \caption{Examples of fits to DQ white dwarfs in our sample. In the
      left panels, error bars represent the observed data, while
      filled circles correspond to our best fit model, with the
      atmospheric parameters given in each panel. The photometry used
      in the fit is indicated at the top left of each panel. A dagger
      symbol indicates that the \logg value has been fixed at 8.0,
      when no trigonometric parallax is available. A star symbol
      indicates that the value of \loghhe has been fixed rather than
      fitted. The right panels show our spectroscopic fits (red) to
      the normalized observed spectra (black). The complete version of
      this figure is available in Appendix I.}
    \label{fig:dqexample}
\end{figure*}

Following the method described in Section \ref{section:method}, we
obtained the atmospheric parameters for all 317 DQ stars in our sample
by fitting simultaneously the spectral energy distribution and the
carbon features (atomic and/or molecular). Figure \ref{fig:dqexample}
shows examples of spectroscopic and photometric fits for typical DQ
white dwarfs (all our fits are available in Appendix I, and our final
parameters are given in Table \ref{tab:resultsdq}). The figures show
the spectral region used for the fit, i.e.~either the Swan bands
between 4000 and 6500 {\AA}, or the carbon lines between 4500 and 5500
{\AA} for the hotter objects. Note that for some objects with weaker
molecular absorption features, we used a smaller region centered on
the observed bands to achieve a good fit.

\begin{figure*}[t]
	\centering
    \includegraphics[width=0.7\textwidth]{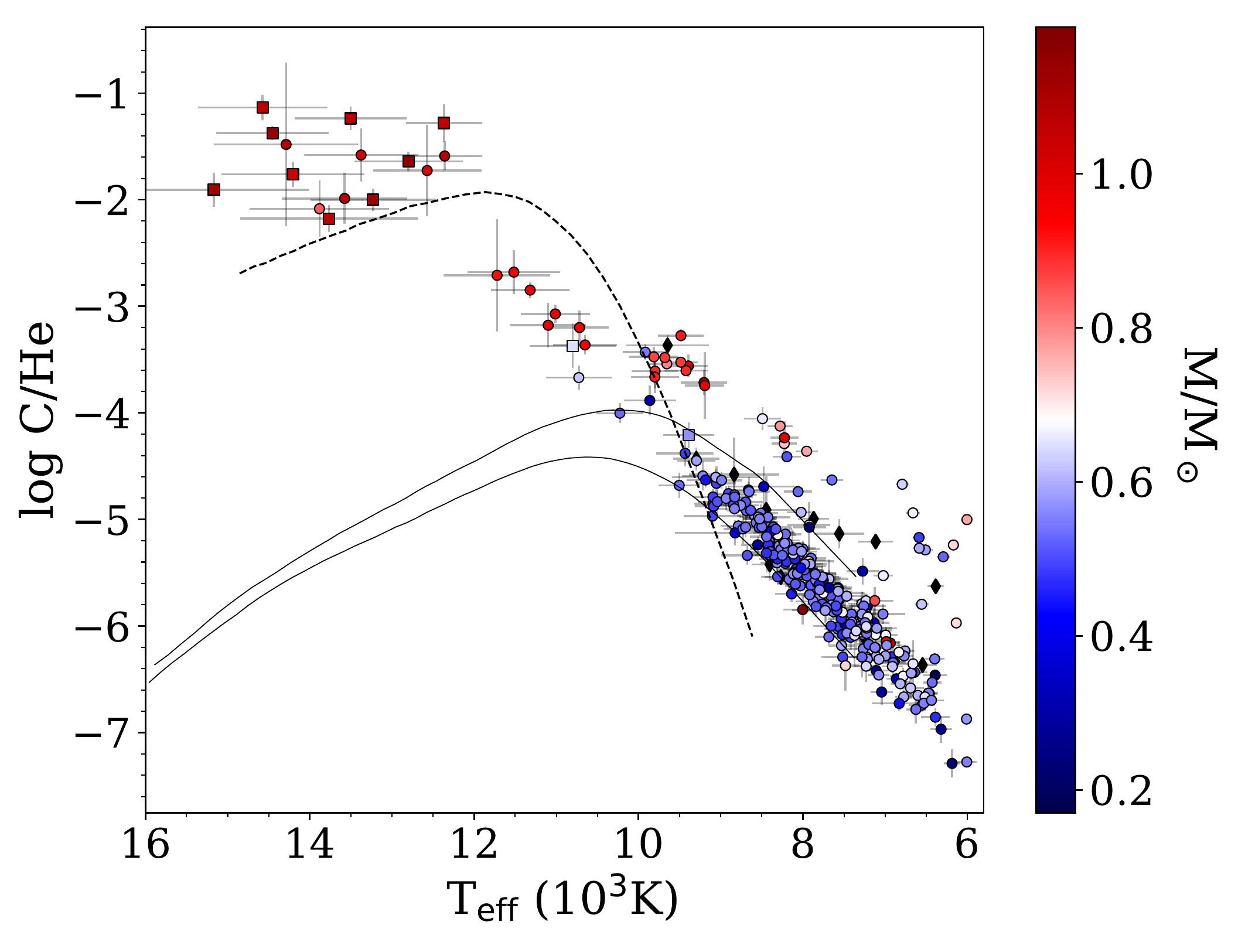}
    \caption{Carbon abundances as a function of effective temperature
      for DQ white dwarfs using a color scale for the mass of each
      object (objects without parallax measurements are shown as black
      diamonds). A square indicates an object fitted with models
      including a trace of hydrogen (\loghhe $=-3$, see text). Solid
      lines represent evolutionary models from \cite{Fontaine2005} at
      0.6 $M_\odot$ with, from top to bottom, $\log q(\mathrm{He})=-3$
      and $\log q(\mathrm{He})=-2$, while the dashed line is for 1.0
      $M_\odot$ and $\log q(\mathrm{He})=-5$. The standard dredge-up
      scenario struggles to explain the observed abundance pattern for
      massive objects.}
  \label{fig:logche}
\end{figure*}

Figure \ref{fig:logche} shows the carbon abundance as a function of
effective temperature, using a color scale to indicate the mass of
each object. Two distinct populations are clearly present: one
sequence with ``normal'' mass ($\sim$0.6 \msun) DQ white dwarfs
(bluish circles) at low effective temperature, and a second sequence
(reddish circles) of massive white dwarfs ($M \ge 0.8$ \msun), with
carbon abundances increasing with effective temperature. The first
sequence follows nicely the expected evolutionary path for 0.6
\msun\ DQ stars with $\log q({\rm He}) = -2.0$ (\citealt{Dantona1979},
\citealt{Iben1985}, \citealt{Fontaine2005}). Note that in
\citet{Dufour2005}, the bulk of the stars was following the
evolutionary sequence for $\log q({\rm He})$ closer to $-3.0$. This is
mainly due to the fact that in our improved models, the band strengths
with our new line list are slightly higher than with the
\citet{Zeidler1982} prescription, resulting in a systematic downward
shift in carbon abundances compared to the values published in
\citet{Dufour2005}. Note also that the abrupt disappearance of normal
mass DQ stars above $\sim$9500 K is simply due to the visibility limit
of carbon at these temperatures; in order to see carbon features in
the optical at higher effective temperatures, the abundance must be at
least \che = $-4.5$. Unless an ultraviolet spectrum is available,
objects following the predicted sequence in that range of effective
temperature will simply be classified as DB white dwarfs. Also, the
absence of stars below \teff $\sim 6500$ K is artificial since these
were not included in our sample due to the uncertainties in modeling
the pressure-shifted Swan bands (the so-called peculiar DQ, or DQpec;
see \citealt{kowalski2010}, \citealt{Blouin2019c}).

More importantly, our analysis confirms the presence of a second
sequence \citep{Dufour2005, KoesterKnist} with an abundance about one
dex higher than the bulk of DQ white dwarfs at $M\sim0.6$ \msun. It
appears that this second sequence is indeed composed of massive white
dwarfs, as suggested by \citet{Dufour2005}, at least for the hottest
ones (\teff $\gtrsim 9000$ K). In fact, almost all DQ white dwarfs
with an effective temperature above 10,000 K have masses higher than
0.8 \msun. The coolest objects on the second sequence, however, do not
appear to have larger masses. These massive DQ stars will be discussed
in greater detail in Section \ref{section:dqmassive}.

\subsection{Mass Distribution} \label{Section:DQmdistr}

\begin{figure}
	\centering
    \includegraphics[width=\columnwidth]{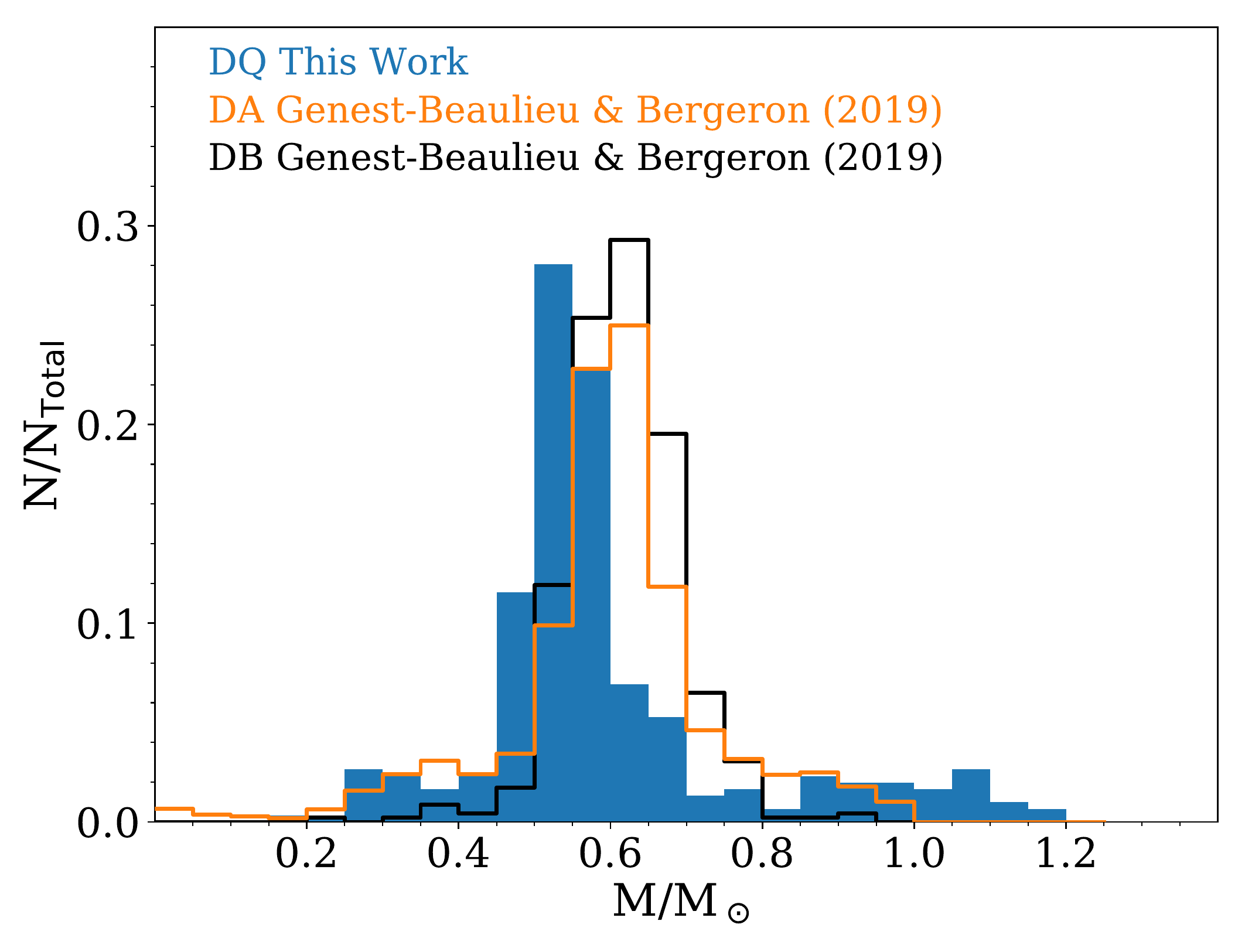}
    \caption{Mass distribution of DQ white dwarfs (blue), compared to
      the photometric mass distributions of DA and DB white dwarfs
      (orange and black lines, respectively) taken from Figure 21 of
      \citet{GenestBeaulieu2019}.}
  \label{fig:histomass_dq_norm}
\end{figure}

The mass distribution of the DQ white dwarfs in our sample, displayed
in Figure \ref{fig:histomass_dq_norm}, reveals the two distinct
populations very clearly, with the bulk of our sample centered around
0.55 \msun, and a second bump centered around 1 \msun\ where all the
hottest DQ stars in our sample are found. As stated before in the case
of DBZ/DZ(A) white dwarfs, DQ stars with derived masses below
$\sim$0.45 \msun\ are probably unresolved double degenerate
binaries. Such low mass stars, if isolated, could not have been formed
from single star evolution within the lifetime of the
Galaxy. Moreover, these objects are usually interpreted as helium-core
white dwarfs whose core mass was truncated by mass transfer with a
companion. The presence of carbon in these apparently low-mass DQ
stars indicates that at least one component has a carbon core. These
binary are thus probably composed of two non-DA stars (e.g., DQ + DC),
which means that the masses reported here are most likely
underestimated.

Surprisingly, the main peak of the mass distribution for the DQ white
dwarfs in our sample is shifted by $\sim$0.05 \msun\ relative to that
obtained for the DA and DB white dwarfs analyzed by \citet[][see their
  Figure 21]{GenestBeaulieu2019}, raising suspicions about our mass
determinations of DQ stars. A way to test the accuracy of our method
for the mass determination is to look at the well-known DQ white dwarf
Procyon B (J0739+0513), which has a very accurate dynamical mass
determination of $0.592 \pm 0.006$ \msun\ \citep{Bond2015}. Using our
standard hybrid photometric/spectroscopic approach, we obtain a
slightly lower mass of $0.554 \pm 0.013$ \msun\ using HST
photometry. This indicates we are probably dealing with a systematic
shift in the mass determinations of DQ white dwarfs using our
models\footnote{Note that a recent study by \citet{Koester2019} also
  finds a DQ mass distribution peaking near 0.55
  \msun.}. Uncertainties related to our dereddening procedure for our
\Te and mass determinations should be irrelevant here given the
proximity of Procyon B. In fact, if we take only objects within 100
pc, where reddening should be minimal along the line of sight for most
stars, we still find that the peak of the mass distribution is too
low. In order to identify the origin of this shift in mass, we
performed several tests using various model grids where we replaced
our treatment of the molecular band opacity with the old just
overlapping line approximation \citep{Zeidler1982}. We also tried
various treatments for the line broadening, and included undetectable
traces of hydrogen or oxygen as well. In the end, similar mass
distributions were always obtained, and we could not pinpoint the
exact cause of this discrepancy.

\begin{figure*}
	\centering
    \includegraphics[width=1.6\columnwidth]{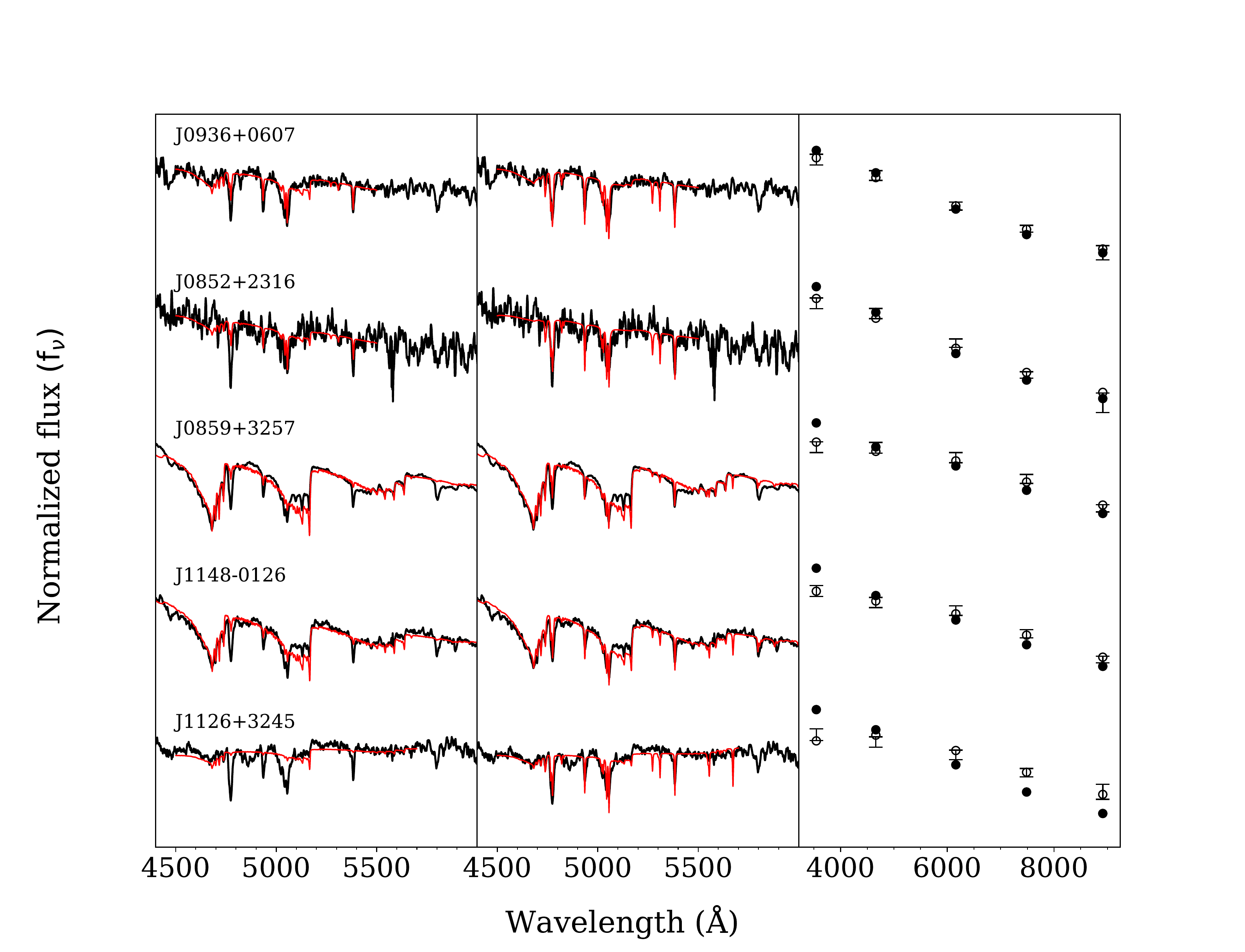}
    \caption{Left panel: best solutions when the effective temperature
      is obtained from fits to photometric data. Middle panel: best
      solutions when fitting only the optical spectra. Right panel:
      corresponding energy distributions for the two solutions. Open
      circles represent the best photometric/spectroscopic solutions
      (our standard approach), while filled circles correspond to the
      solutions with the effective temperature and carbon abundance
      determined solely from the optical spectra.}
  \label{fig:spectrofit}
\end{figure*}

We note, however, that several lines in the ultraviolet have wings
that extend several hundreds of angstroms from the line center. This
is probably not physically realistic, and the use of the impact
approximation is certainly not appropriate for these
lines. Unfortunately, the data required to compute line profiles with
the unified line shape theory of \citet{Allard1999} are not available
for these lines. For stars with available spectroscopic observations
in the UV
\citep{Holberg2003}\footnote{\url{http://vega.lpl.arizona.edu/newsips/low/}},
the use of the quasistatic van der Waals broadening
\citep[][D.~Koester, private communication]{Walkup1984} provides a
good fit to the asymmetric C \textsc{i} 1930 {\AA} line, although the
carbon abundance needed to reproduce the observations is different
than that determined from the Swan bands. This problem is not new
\citep[see][for a detailed discussion]{Dufour2011b}, and the origin of
this discrepancy still remains mysterious to this day, although it is
most likely related to the uncertainties of the ultraviolet
opacities. In fact, for DQ stars in the intermediate temperature
regime where both atomic lines and molecular bands are present
($\sim$9500 to 11,000~K), the models also have difficulties
reproducing both types of absorption features simultaneously. Examples
of this problem are illustrated in Figure \ref{fig:spectrofit}. While
it is possible to obtain a good spectroscopic fit by increasing the
effective temperature and carbon abundance for these objects, doing so
would then produce an energy distribution completely at odds with the
photometric observations (see right panel of Figure
\ref{fig:spectrofit}). A similar problem, based on an analysis using
D.~Koester's model atmosphere code, was also reported by
\citet{Gansicke2010} for two DQ stars showing traces of oxygen. For
these problematic DQ stars in our sample, we compare in Table
\ref{tab:badspecfit} the results of our standard approach with the
parameters obtained by fitting only the optical spectra, thus ignoring
all photometric information. We note that even though the exact
atmospheric parameters are somewhat uncertain because of this problem,
the massive nature ($M > 0.8$ \msun) of these objects remains
unquestionable, even when allowing for a conservative uncertainty of
0.15 \msun.

While the problems mentioned above become apparent only for stars that
show both atomic and molecular lines, it is likely, however, that the
whole temperature, abundance, and mass scales for DQ white dwarfs are
affected, which could perhaps also explain the $\sim$0.05 \msun\ shift
of the peak of the DQ mass distribution relative to DA and DB white
dwarfs. Until this problem is solved, the absolute values of the
atmospheric parameters for all DQ white dwarfs should be considered
uncertain, although the relative values between objects in the sample
are probably reliable.

\begin{table*}
\centering
\caption{Parameters of DQ white dwarfs with bad spectroscopic fits when the effective temperature is derived from photometry}
\begin{tabular}{c|cccccc}
\hline
&\multicolumn{3}{c}{Photometric fits}&\multicolumn{3}{c}{Spectroscopic fits}\\
Name& $T_\mathrm{eff}$ (K)&   $\log{\textrm{C}/\textrm{He}}$ & $M$/\msun  & $T_\mathrm{eff}$ (K)&   $\log{\textrm{C}/\textrm{He}}$ &  $M$/\msun\\
\hline
J0852+2316 &11099&$-$3.18&0.97 &12733&$-$1.99 &1.12\\
J0859+3257            &9486 &$-$3.52&0.87 &10407&$-$2.87 &0.97\\
J0936+0607 &11013&$-$3.07&0.97 &12109&$-$1.99 &1.09\\
J1000+1005 &7869 &$-$4.99& &10395&$-$2.87 &\\
J1126+3245 &9297 &$-$4.43& &11369&$-$2.73 &\\
J1140+0735 &10651&$-$3.36&0.94 &12395&$-$1.60 &1.12\\
J1140+1824 &9656 &$-$3.54&0.81 &10921&$-$2.62 &0.94\\
J1148$-$0126         &9680 &$-$3.48&0.88 &10868&$-$2.62 &1.00\\
J2248+2826 &9390 &$-$4.21&0.57 &11920&$-$2.02 &0.90\\
\hline
\end{tabular}
\label{tab:badspecfit}
\end{table*}

\subsection{The Effects of Hydrogen} \label{section:dqhydro}

The presence of hydrogen is a rare phenomenon in DQ white dwarfs
\citep[see][]{Dufour2011b}. Hydrogen abundance measurements are
reported only for two relatively hot and massive DQ stars in the
literature, namely G35-26 \citep{Thejll1990} and G227-5
\citep{Wegner1985}, while its presence is also inferred from the
detection of a CH molecular band in two other objects --- G99-37 and
GJ 841B, both magnetic white dwarfs
\citep{Blouin2019c,Vornanen2010}. Our sample of cool DQ white dwarfs
with molecular carbon bands contains only six objects showing
H$\alpha$, three of which have masses below 0.47 \msun, indicating
they are most probably unresolved double degenerate binaries composed
of a DQ and a DA white dwarf; these are listed in Table
\ref{tab:DQAlist}. \citet{Leggett2018} also identified two such DQ+DA
unresolved binaries (see their Figure 7). Except for the hotter DQ
J1243+1651, the other objects have an \halpha line that is too narrow
to be reproduced by helium-rich models, and we believe they are also
DQ + DA binaries. Hence, of the 293 DQ white dwarfs with temperatures
between $\sim$6500 K and 10,500 K in our sample, there is compelling
evidence for the presence of hydrogen in only two objects, both hotter
than 9000 K.

\begin{table}[h!]
\centering
\caption{DQ white dwarfs showing hydrogen lines and Swan bands}
\begin{tabular}{c|cc}
\hline
Name & \teff (K) & M/\msun \\
\hline
J0836+0437 & 8420 & 0.46 \\ 
J0928+2638 & 7108 & 0.28 \\
J0950+3238 & 8268 &      \\
J1243+1651 &10227 & 0.53  \\
J1406+3402 & 7042 & 0.30 \\
J2310$-$0057 & 7647 & 0.54 \\
\hline
\end{tabular}
\label{tab:DQAlist}
\end{table}

As mentioned in Section \ref{section:dzfits}, undetectable traces of
hydrogen can still have an impact on the atmospheric parameter
determination of helium-atmosphere white dwarfs, although in the case
of DQ white dwarfs, the contribution of free electrons from carbon is
usually much more important than that of heavy elements in DZ
stars. To test the effect of hydrogen on the determination of the
atmospheric parameters of cool DQ white dwarfs, we generated a small
grid of models with \loghhe $=-3$, and refitted every object in our
sample with \teff $<10,000$ K. We find that by assuming such a
hydrogen abundance, we could produce H$\alpha$ and H$\beta$ absorption
features that would be easily detected above 8000 K, as well as a CH
molecular band near 4300 {\AA} for objects cooler than $\sim$8500
K. Even by assuming an abundance that is clearly ruled out by the
spectroscopic observations, we find that the presence of hydrogen has
only a marginal impact on our atmospheric parameters, increasing the
derived effective temperatures by 150 K and masses by 0.035 \msun, on
average. We can thus safely consider that the presence of undetectable
traces of hydrogen does not affect significantly the parameters of
cool DQ stars.

In the atomic lines regime (\teff $\gtrsim 10,000$ K), however,
spectroscopic observations tell a different story. Although H$\alpha$
is blended with the carbon lines near 6588 {\AA}, H$\beta$ can be
detected very clearly. These lines are observed in 40\% of the objects
(10/25). We thus generated additional grids with \loghhe $=-2$, $-3$,
and $-4$, with \teff ranging from 10,000 K to 16,000K, We found that,
surprisingly, the \loghhe $=-3$ grid best reproduces the observations
for almost every star, suggesting that the hydrogen abundance appears
relatively constant in the objects where hydrogen is
detected. Therefore, the solutions with \loghhe $=-3$ are presented in
Table \ref{tab:resultsdq} when the Balmer lines are observed. Note
that it is possible that DQ stars with no detectable hydrogen features
also contain some traces of hydrogen, since the Balmer lines are
barely predicted for some objects with higher carbon abundance and
lower effective temperature. The presence of hydrogen in these objects
has a marginal impact on the measured effective temperature (265 K on
average) and mass (0.032 \msun\ on average), but the effect can be
quite significant for the carbon abundance (around 0.5 dex in some
cases). Since we do not know the exact hydrogen abundance for these
objects, we provide the solutions without hydrogen in Table
\ref{tab:resultsdq}.

\subsection{Massive DQ White Dwarfs} \label{section:dqmassive}

We now turn our attention to the population of massive ($M > 0.8$
\msun) DQ white dwarfs clearly distinguishable in Figures
\ref{fig:logche} and \ref{fig:histomass_dq_norm}. First, the carbon
abundance pattern observed for these objects indicates that they
followed a different evolutionary path than their cooler normal mass
counterparts, since the standard dredge-up prediction for massive
white dwarfs completely fails to explain their chemical composition
\citep[see also][]{brassard2007}. Second, such massive DQ stars could
not have DB white dwarfs as progenitor because practically no massive
helium-rich white dwarfs exist at higher effective temperatures
\citep{Bergeron2011, GenestBeaulieu2019, Beauchamp1996}. Instead, the
massive sequence connects nicely with the carbon-dominated atmosphere
white dwarfs at higher effective temperature, the so-called Hot DQ
stars \citep[\teff $\ge$ 18,000~K and \che $\ge$ 0,
  see][]{Dufour2007Nat, Dufour2008, Dufour2013}. Note that {\it Gaia}
trigonometric parallax measurements confirm that the carbon-dominated
atmosphere white dwarfs are massive as well (Dunlap et al.,
submitted). We thus believe that our massive DQ stars represent cooler
versions of the carbon-atmosphere white dwarfs.

The many unusual properties of Hot DQ white dwarfs --- unique chemical
composition, high mass, and high incidence of magnetism --- recently
prompted \citet{Dunlap2015} to propose that these hot
carbon-atmosphere white dwarfs represent a population of merged white
dwarfs (failed type Ia supernovae). The key piece of evidence for this
proposed scenario comes from the kinematic properties of the sample,
which provide an independent age indicator \citep{Dunlap2015}. Indeed,
one interesting characteristic that stands out about the Hot DQ
population is their very high tangential velocity, as a group,
compared to other white dwarfs with the same age and mass. Over time,
a population of stars is kinematically heated through gravitational
interactions \citep{Wegg2012}. Hence, as a population gets older, its
velocity dispersion increases. In particular, Dunlap \& Clemens showed
that there is a discrepancy between the young age inferred from the
derived atmospheric parameters of carbon-atmosphere white
dwarfs\footnote{Such hot massive white dwarfs should descend from
  massive, short-live main-sequence progenitors, and they should thus
  have a small total age if they evolved as single stars.}, and the
old age derived from the velocity dispersion. The merger scenario
provides an elegant solution to this dilemma because the reheating
induced by the merging event resets the cooling age clock, and
consequently, the cooling age derived from the effective temperature
and mass thus becomes meaningless. If the massive DQ white dwarfs
observed in Figures \ref{fig:logche} and \ref{fig:histomass_dq_norm}
are indeed cooled down versions of the Hot DQ stars, a kinematic
analysis of the sample should then show the same discrepancy between
the two age indicators.

\begin{figure}[t]
	\centering
    \includegraphics[width=\columnwidth]{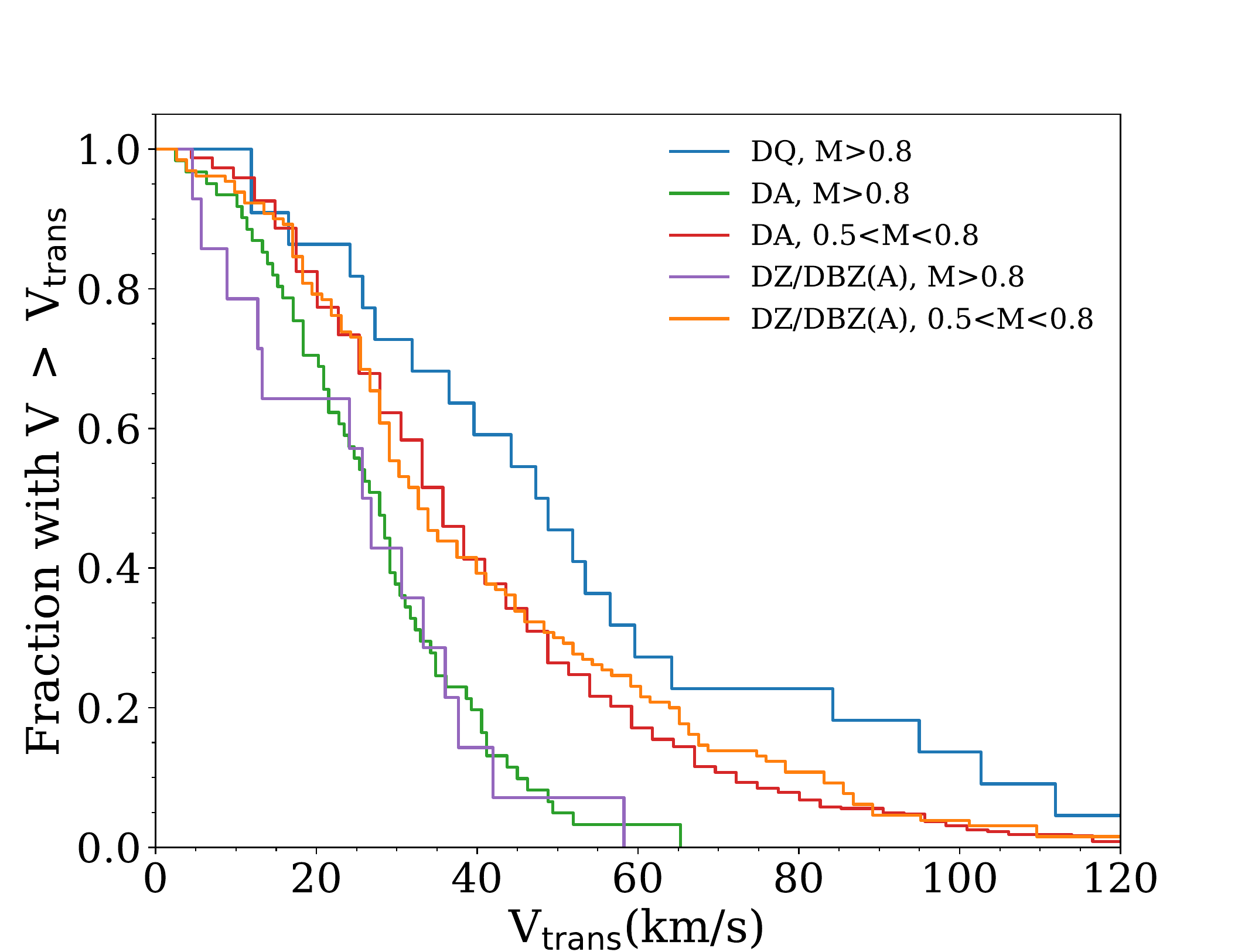}
    \caption{Cumulative distribution of transverse velocities for
      massive DQ, DBZ/DZ(A) (this paper) and DA
      \citep{GenestBeaulieu2019} white dwarfs with \teff between
      10,000 and 16,000K.}
    \label{fig:vtrans}
\end{figure}

We show in Figure \ref{fig:vtrans} the cumulative distribution of
transverse velocities, $V_{\rm trans}$, for our sample of massive DQ
white dwarfs, calculated using {\it Gaia} distances and proper
motions. We selected DQ white dwarfs with \Te $> 10,000$~K and $M>0.8$
\msun, a point where there appears to be a clean separation between
the two DQ populations. Since there is no DQ white dwarf with
``normal'' masses in this temperature range, we compare the properties
of the massive DQ stars with those of our sample of DBZ/DZ(A) as well
as DA white dwarfs with signal-to-noise ratio higher than 25 taken
from \citet{GenestBeaulieu2019}, a large and clean sample perfectly
suited for this comparison. Also, in order to make a meaningful
comparison, we selected only DBZ/DZ(A) and DA white dwarfs in the same
range of effective temperature where the massive DQ stars are found,
i.e.~between 10,000 K and 16,000 K, and we compare the distributions of
transverse velocities for stars with masses greater than 0.8 \msun\ and
in the $0.5-0.8$ \msun\ mass range.

First, we notice that the transverse velocity distribution for
DBZ/DZ(A) and DA white dwarfs are very similar, and that those with
$M>0.8$ \msun\ are much less dispersed than those with $0.5 < M < 0.8$
\msun. This is exactly what we expect, since massive white dwarfs have
massive progenitors with shorter main sequence lifetimes,
corresponding to a shorter total age, and thus a smaller velocity
dispersion \citep{Wegg2012}. However, the velocity dispersion for the
massive DQ stars displayed in Figure \ref{fig:vtrans} is extremely
broad, in sharp contrast with what is expected for a population of
massive white dwarfs. We find indeed that 45\% (10 out of 22) of the
massive ($M > 0.8$ \msun) DQ population has $V_{\rm trans} > 50$ km
s$^{-1}$, while the fractions are only 5\% (3 out of 61) and 7\% (1
out of 14) for DA and DBZ/DZ(A) with similar atmospheric parameters,
respectively. The fact that the distribution is even larger than that
of stars in the $0.5-0.8$ \msun\ mass range strongly supports the idea
that these objects are the descendants of the carbon-dominated
atmosphere white dwarfs (the Hot DQ stars), and consequently, that
they represent the outcome of the merging of two white dwarfs.

It is interesting to note, however, that very few massive DQ stars are
found with a large magnetic field (J1040+0635 and J1036+6522 are the
only two examples, \citealt{Williams2013}), in contrast with the very
large fraction of magnetic Hot DQ white dwarfs \citep[at least 70\%
  are magnetic at some level,][]{Dufour2013}. Magnetic fields of a few
MG would be easily detectable through line splitting in our
spectra. It thus seems that as helium finds its way to the top of the
photosphere, the strength of the magnetic field is also reduced
significantly, as these stars cool down. If our hypothesis that the
Hot DQ stars represent the progenitors of the massive DQ in our sample
is correct, then the latter are probably still magnetic at some
level. Unfortunately, the SDSS spectra used in our analysis lack the
signal-to-noise ratio and spectral resolution required to properly
identify weaker magnetic fields of a few 100 kG. We predict that a
large fraction of these objects will eventually show magnetism when
high-resolution spectroscopic observations become available. Also, at
least one third of the Hot DQ stars are variable
\citep{Montgomery2008, Barlow2008, Dunlap2010, Lawrie2013,
  Dufour2011}, presumably due to the presence of magnetic spots at the
surface of a rapidly rotating star. If magnetism is indeed still
present at some level, a large fraction of the massive DQ stars should
also be variable.

Finally, we notice that there appears to be a correlation between the
mass and the effective temperature, the hottest objects near 16,000 K
being significantly more massive ($\sim$0.3 \msun) than their cooler
counterparts near 10,000 K. As it is expected that white dwarfs should
evolve at constant masses, we first believed that some missing
ingredient in our models was responsible for either an overestimation
of the masses of the hottest stars, or an underestimation for the
cooler ones. We experimented with many test grids incorporating traces
of oxygen, hydrogen, as well as different treatments of line
broadening, but none of these experiments affected the relative masses
of our object in a significant way. As mentioned in Section
\ref{Section:DQmdistr}, we know there are some uncertainties
concerning the absolute values of the atmospheric parameters of DQ
white dwarfs that may explain the shifted position of the peak of the
mass distribution and the difficulties in reproducing simultaneously
the atomic and molecular features. However, it is unlikely that the
shortcomings of our models translate into uncertainties on the masses
much larger than 0.1 \msun\ (the radius of the star is tightly
constrained from the measured energy distribution and distance), not
enough to explain the clear correlation with effective temperature
that we observe in our sample. We thus believe that the observed
correlation is real and that it is most probably the manifestation of
an observational bias due to the crystallization of the core of these
white dwarfs (see next section).

\subsection{The Crystallization Sequence}

As a white dwarf cools off, thermal energy is gradually lost from the
star in the form of radiation, until the kinetic motions of the ions
lose amplitude and eventually become correlated. The ionic state then
evolves from a gas to a fluid to a solid, a process referred to as
crystallization \citep{Fontaine2001}. This liquid-to-solid transition,
which begins in the center of the stellar core and slowly moves
outwards, represents a first-order phase transition, and is thus
accompanied by a release of latent heat, which contributes to slowing
down the cooling process significantly. Over 50 years ago,
\cite{vanHorn1968} predicted that this crystallization process would
cause a decrease of the cooling rate during the transition, and would
cause a pile-up of objects that could be detected in HR diagrams. The
first direct observational evidence of this crystallization process
was reported by \cite{Tremblay2019}, in the form of a characteristic
pile-up of white dwarfs forming a tight sequence in the {\it Gaia}
M$_G$ versus $G_\mathrm{BP}-G_\mathrm{RP}$ HR diagram, the so-called
``Q branch'' (see their Figure 2).

Our second, more massive sequence of DQ white dwarfs is particularly
interesting in the context of crystallization, because for $\sim$0.6
\msun\ white dwarfs, crystallization occurs at the same time as
another physical process referred to as convective coupling, which
takes place when the base of the superficial convection zone reaches
into the degenerate interior \citep{Fontaine2001}. Convective coupling
also decreases the cooling process temporarily, and in $\sim$0.6
\msun\ white dwarfs, the effects of crystallization and convective
coupling cannot be disentangled. Fortunately, crystallization occurs
at much higher effective temperature for more massive white dwarfs,
making the massive sequence of DQ white dwarfs ideal objects to
examine the crystallization process.

\begin{figure}[t]
	\centering
    \includegraphics[width=\columnwidth]{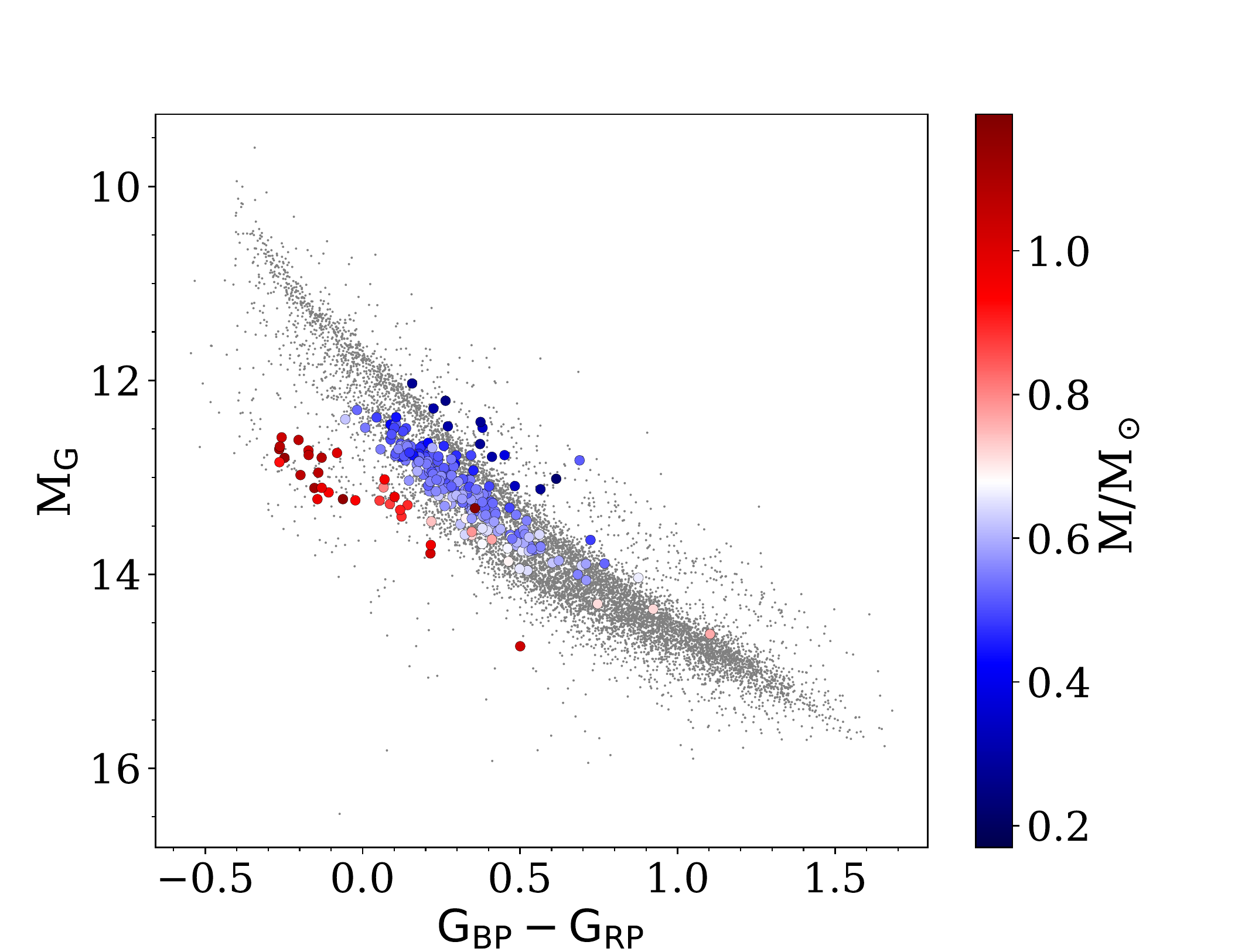}
    \caption{Absolute $G$ magnitudes as a function of
      $G_\mathrm{BP}-G_\mathrm{RP}$. Filled circles represent DQ stars
      with $\sigma_\pi / \pi < 0.1$, using a color scale for the mass
      of each object. Black dots are white dwarf candidates within 100
      pc selected from {\it Gaia} DR2 using the cuts proposed in
      Appendix B of \cite{Gaia2018}.}
    \label{fig:colmag_dq}
\end{figure}

Figure \ref{fig:colmag_dq} shows the color-magnitude diagram for our
sample of DQ white dwarfs with measured parallaxes and $\sigma_\pi /
\pi < 0.1$. The lowest red point in this diagram is J0841+3329, for
which the {\it Gaia} \texttt{phot\_bp\_rp\_excess\_factor} value is
1.34, indicating possibly a bad quality of magnitudes according to
\cite{Gaia2018}. Our best fit based on SDSS photometry predicts a
lower magnitude in the $G$ bandpass. By removing this object, all our
massive DQ stars fall in the region of the crystallization sequence
reported by \cite{Tremblay2019}.

\begin{figure*}[t]
	\centering
    \includegraphics[width=\textwidth]{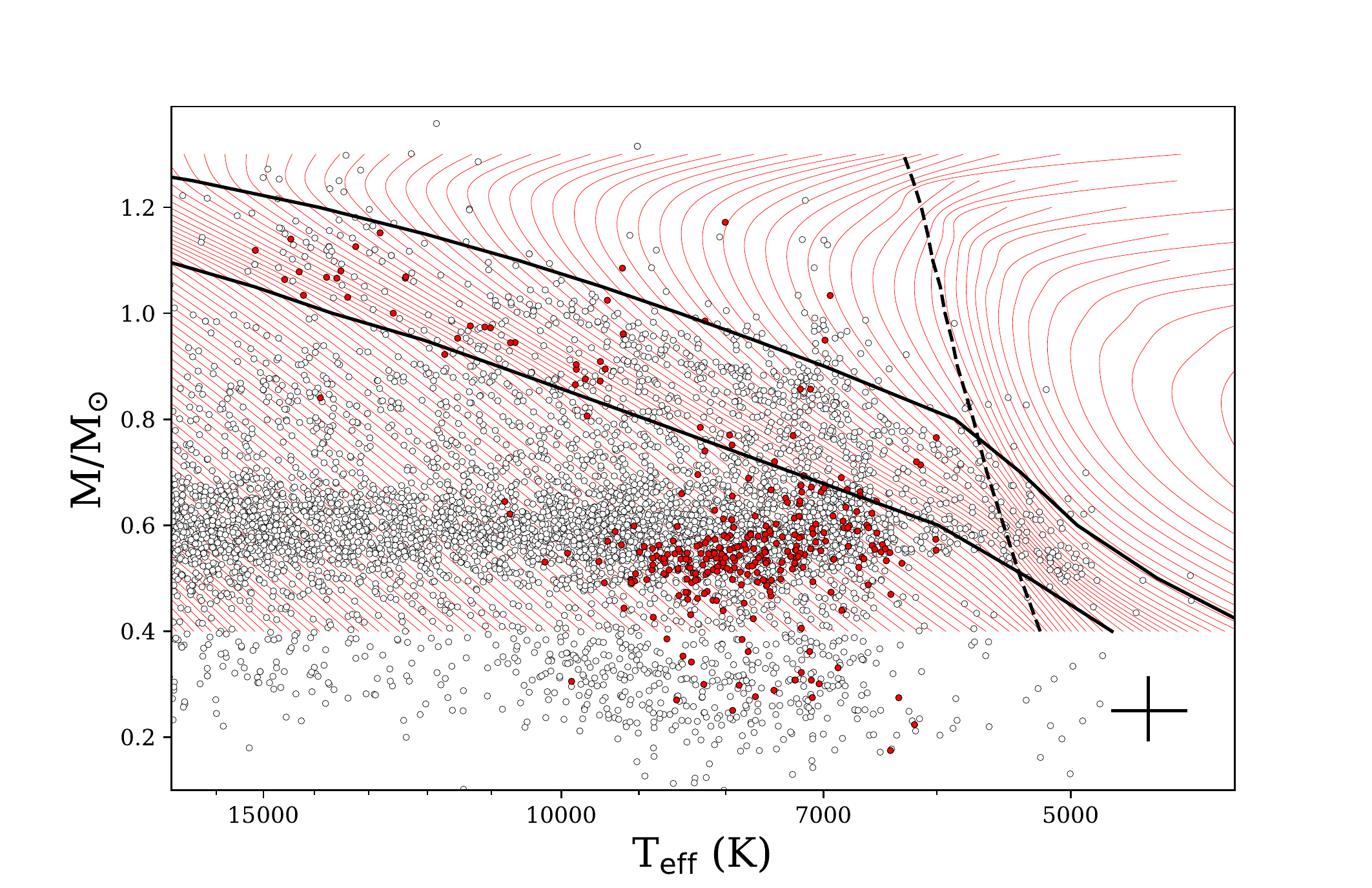}
    \caption{Mass as a function of \teff for the DQ white dwarfs in
      our sample (red dots), and DA white dwarfs spectroscopically
      identified in the MWDD with $\sigma_\pi/\pi < 0.1$ (white dots),
      taken from Figure 14 of \citet{Bergeron2019}. Also reproduced
      are their theoretical isochrones obtained from cooling sequences
      with C/O-core compositions, $q({\rm He})=10^{-2}$, and $q({\rm
        H})=10^{-4}$, equally spaced by $\Delta
      \log{\tau_\mathrm{cool}}=0.02$ (in years). The lower black solid
      curve indicates the onset of crystallization at the center of
      evolving models, while the upper one indicates the location
      where 80\% of the total mass has solidified. The dashed curve
      indicates the onset of convective coupling. The black cross
      corresponds to the mean errors of each fitted parameter.}
  \label{fig:massteff}
\end{figure*}

Figure \ref{fig:massteff} shows the masses as a function of effective
temperature for the same sample of DQ stars, together with the sample
of DA white dwarfs analyzed by \citet[][see their Figure
  14]{Bergeron2019}. The black solid lines are also described at
length in Bergeron et al. Briefly, the left line corresponds to the
onset of crystallization at the center of an evolving model \cite[the
  same as in][]{Tremblay2019}, while the right line indicates the
location where 80\% of the total mass of the star has solidified. Upon
crystallization, latent heat is slowly released, and the white dwarf
cooling is slowed down, a process that is well illustrated by the
tightening of the isochrones (shown in red) between the two black
lines. Figure \ref{fig:massteff} reveals that most of our massive DQ
stars fall within these lines, indicating a possible pile-up due to
crystallization.

\begin{figure*}
	\centering
    \includegraphics[width=\textwidth]{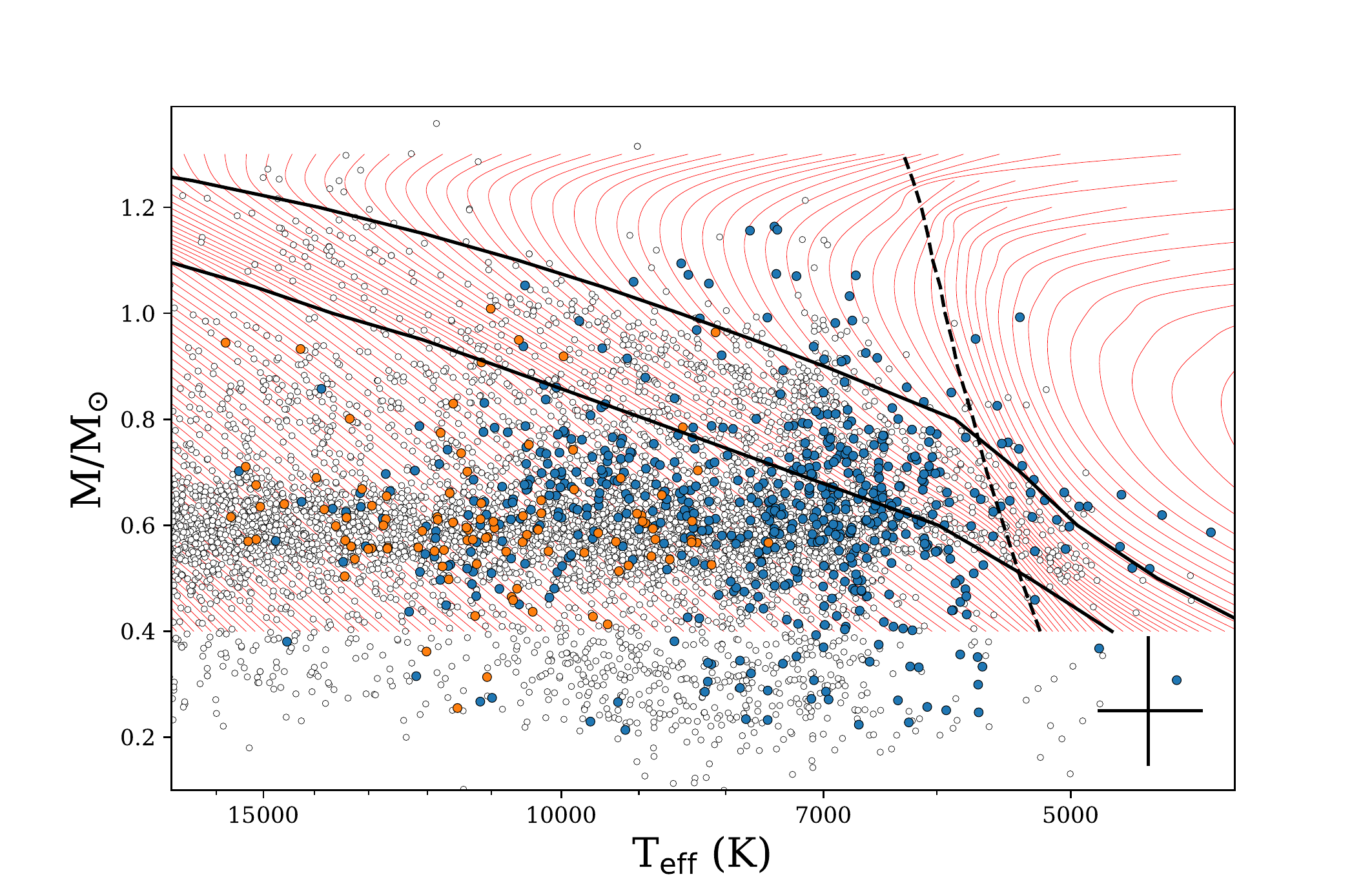}
    \caption{Same as Figure \ref{fig:massteff} but for DZ/DBZ white
      dwarfs. Blue circles correspond to DZ/DBZ stars, while orange
      circles represent DZA/DBZA stars.}
  \label{fig:massteff_dz}
\end{figure*}

Figure \ref{fig:massteff_dz} shows the masses of DBZ/DZ(A) white
dwarfs in our sample as a function of effective temperature together
with the same isochrones and crystallization sequences as before. The
signature of crystallization cannot be seen, but this could be due to
the fact that there are much less massive stars above \teff $=$ 12,000
K which is consistent with the idea that DBZ(A) white dwarfs belong to
the same population as DB(A) stars. At high temperatures, it is also
more difficult to see metal absorption lines (see the detection limit
in Figure \ref{fig:logcahe}) and the diagram is thus much less populated
in that range.

\subsection{Discussion of Individual Objects}

\textit{J0901+5751}, \textit{J0922+2928}, and \textit{J1423+5729} ---
These stars show many O \textsc{i} lines, especially at 6156 and 7772
{\AA}, and at 5330 and 5435 {\AA} in the case of J1423+5729. For
J0922+2928, \cite{Gansicke2010} found \teff $=8270$ K and \logche
$=-2.6$ using oxygen-rich models, while we found \teff $=8022$ K and
\logche $=-4.93$. They also found \loghhe $<-5.0$ and
$\log{\mathrm{O}/\mathrm{He}} = -2.0$, which implies that oxygen is
more abundant than carbon. White dwarfs with such a high amount of
oxygen in their atmosphere must have followed a different evolutionary
path, and the authors suggest that they could be O/Ne-core white
dwarfs surrounded by a layer of carbon and oxygen. Such white dwarfs
would be massive. According to our results, only one of these can be
qualified as massive, but we determined the mass using evolutionary
models with C/O cores and did not take atmospheric oxygen into account
in our analysis. However, if we take the values of \cite{Gansicke2010}
for J0922+2928, all these objects fall above the first sequence in
Figure \ref{fig:logche}, suggesting indeed a different evolutionary
path. We also note that these objects are rare and have little impact
on the conclusions of this work.
	
\textit{J1040+0635} --- This star shows magnetic splitting, suggesting
that our solution may be uncertain. To evaluate the influence of the
magnetic field on our result, we fitted the magnetic white dwarf
J1036+6522 (not included in this analysis), which shows similar line
splitting. We found \teff $= 15,637$ K, \logg $= 8.8$, and \logche
$=-1.87$, the carbon abundance being very approximate since we do not
include splitting in this exercise. \cite{Williams2013} used magnetic
synthetic spectra to fit this object and estimated $T_\mathrm{eff}\sim
15,500$ K, $\log{g} \sim 9$, and $\log{\mathrm{C}/\mathrm{He}}=-1$,
which is consistent with our solution.

\textit{J2101+3148} --- This star shows a discrepancy for the Swan
band at 4700 {\AA}. \cite{Dufour2005} also reported this problem
without further explanation. With a larger sample, we can now compare
this object with other stars with similar properties (for example
J1424+0833 and J1118$-$0314), and this discrepancy is not observed
anywhere else. A change in temperature of $\pm 500$ K does not improve
the situation.  However, the finding chart on SIMBAD clearly shows a
red object very near the white dwarf (potentially a companion or an
object along the line of sight) that probably contaminates the
spectra.

\section{On The Spectral Evolution of White Dwarfs} \label{Section:spectralevolution}

Thanks to the advent of {\it Gaia}, we now have, for the first time,
detailed mass distributions for a large sample of DQ and DBZ/DZ(A)
white dwarfs. Although there are still uncertainties regarding the
individual mass determinations (see Sections \ref{section:dzfits} and
\ref{Section:DQmdistr}), a lot of information can be extracted from the
overall shape of these mass distributions, and in particular with
respect to the spectral evolution of these white dwarfs.

The mass distribution of DQ white dwarfs shows two distinct peaks (see
Figure \ref{fig:histomass_dq_norm}), one centered around 0.55
\msun\ and another one centered around 1 \msun. As discussed above,
the location of the first peak is slightly shifted relative to that of
DA and DB white dwarfs \citep[see Figure 21 of][]{GenestBeaulieu2019},
most probably indicating a systematic error due to some unknown
opacity source in the ultraviolet. Putting aside this problem, it is
interesting to compare the overall shape of these mass
distributions. The main difference between the mass distribution of DB
stars and that of DA stars is the absence of a high-mass tail in the
former. This difference was first noted by \citet{Beauchamp1996}, and
later confirmed by \citet{Bergeron2011} and
\citet[][]{GenestBeaulieu2019}. The mass distribution of our DQ sample
does not show an extended tail extending to masses above 1 \msun, but
rather exhibits two distinct peaks with a clear separation near 0.8
\msun. As discussed in Section \ref{section:dqmassive}, there is
compelling evidence that the second peak represents a population of
merged white dwarfs. Hence, the mass distribution of DQ stars with
\teff < 10,000 K --- effectively removing practically all the massive
objects --- resemble much more that of DB white dwarfs than that of DA
stars, reinforcing the belief that the progenitor of DQ white dwarfs
are helium-rich DB stars.

\begin{figure}
	\centering
    \includegraphics[width=\columnwidth]{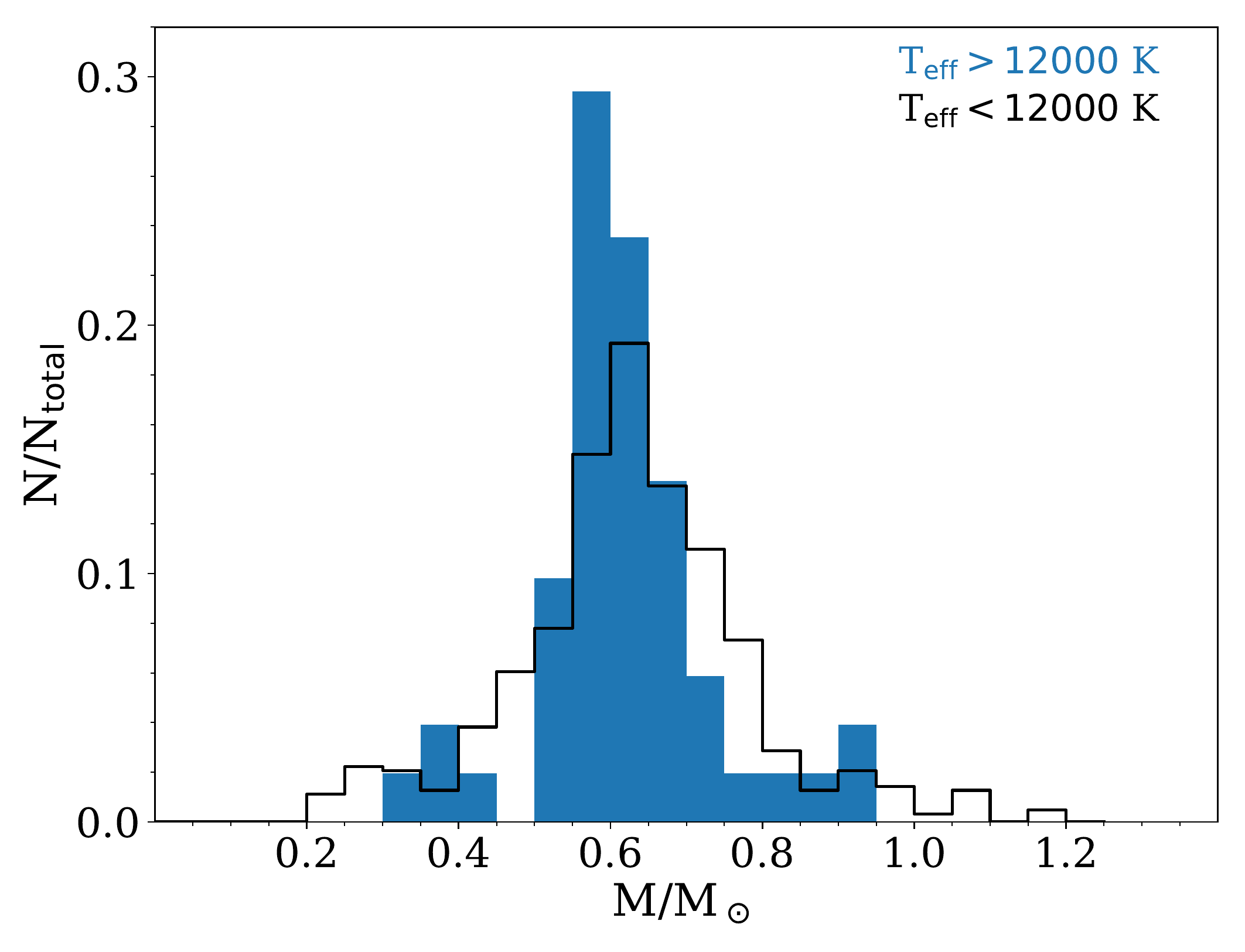}
    \caption{Comparison of the mass distributions of DBZ/DZ(A) white
      dwarfs for effective temperatures above and below 12,000 K.}
  \label{fig:evolv}
\end{figure}

In contrast, the mass distribution of DBZ/DZ(A) white dwarfs (see
Figure \ref{fig:histomass}) reveals a fair number of massive objects
forming some sort of a tail similar to that observed in DA white
dwarfs. The similitude is even more striking in Figure \ref{fig:evolv}
where we show the mass distributions of our sample for effective
temperatures above and below 12,000 K. We see that at low effective
temperatures, the mass distribution --- composed essentially of DZ(A)
stars --- clearly shows a high mass tail, which is barely present in
the mass distribution of hotter white dwarfs, containing essentially
only DBZ(A) white dwarfs.

The most logical explanation for this phenomenon is that the
progenitors of a significant fraction of DZ(A) white dwarfs are DA
stars that were convectively mixed, and transformed into
helium-atmosphere white dwarfs. Indeed, it is well known that the
ratio of non-DA to DA white dwarfs increases considerably with
decreasing effective temperatures. This has been interpreted as the
result of convective mixing, a process which becomes important below
12,000 K, as discussed at length in \cite{Rolland2018} and references
therein. This is a process where the thin hydrogen layer is mixed with
the deeper and more massive helium layer through convection,
effectively turning a hydrogen-rich white dwarf into a helium-rich
one. Our results thus suggest that many DZ(A) stars, and certainly
most of the massive ones, represent the outcome of mixed DA white
dwarfs that have also accreted planetary material. In fact,
\cite{Rolland2018} also demonstrated that most DZA stars below \teff
$\sim$ 12,000 K could not be the descendants of DBA (or DBZA) stars,
and that these objects must be the result of hydrogen-rich stars that
turned helium-rich as a result of convective mixing. For the DBZ(A)
white dwarfs, it is reasonable to consider these objects as being
simply the metal-polluted version of DB(A) stars. These polluted
DBZ(A) stars, when they cool off, will eventually become helium-rich
white dwarfs, or DZ(A) stars if they also accrete material (they would
not become DQ stars, however, see below).

Finally, the DQ white dwarfs with masses near the peak of the mass
distribution --- which rarely show any trace of hydrogen (see Section
\ref{section:dqhydro}) --- are most likely the descendants of the
so-called ``pure helium'' DB stars that show no hydrogen feature. In
these nearly hydrogen-free DB stars, hydrogen has probably been
largely depleted during the earlier born-again post-AGB evolutionary
phase \citep{Rolland2018}. This interpretation could simultaneously
explain the rarity of DQZ objects\footnote{There are only 4 stars with
  Ca \textsc{ii} H\&K absorption lines in our sample: J0739+0513
  (Procyon B), J0900+0331, J1332+2740, and J1534+4145.} (i.e. DQ stars
showing traces of metals), if the surroundings of the progenitor have
been cleared of rocky debris during the active earlier post-AGB
phases.

\section{Summary and Conclusions} \label{Section:conclu}

We presented a homogeneous analysis of 1023 DBZ/DZ(A) and 319 DQ white
dwarfs based on state-of-the-art model atmospheres using new
parallaxes, proper motions, and photometry from {\it Gaia} DR2, as
well as photometry from SDSS, Pan-STARRS, previously published
\textit{BVRI}, and spectroscopy from various sources. This represents a
significant increase over the previous comprehensive studies on these
types of objects, namely those of \citet[][159 DZs]{Dufour2007} and
\citet[][56 DQs]{Dufour2005}. Calcium abundance measurements for our
large sample of DBZ/DZ(A) white dwarfs indicate that the rocky objects
that polluted their photosphere had masses similar to those of large
asteroids found in our solar system. We found several polluted white
dwarfs with progenitor masses well above 3 \msun, confirming that the
formation of rocky material is also common for early type stars. The
availability of parallax measurements for nearly three quarters of our
sample allowed us to determine, for the first time, meaningful mass
distributions for these types of objects (the mass distributions for
DQ and DZ stars of \citealt{Dufour2005, Dufour2007} contained only 11
and 16 objects, respectively). These mass distributions revealed
several interesting aspects about the properties of our samples that
we summarize here:

\begin{enumerate}

\item The mean mass for the DBZ/DZ white dwarfs (i.e. objects not
  showing H$\alpha$) is significantly higher than that of DBZA/DZA
  stars. The two distributions are in much better agreement when
  undetectable traces of hydrogen are included in the model fits.

\item The mass distribution of DZ(A) white dwarfs cooler than $T_{\rm
  eff}=12,000$ K shows a high-mass tail similar to that observed for
  DA stars. This high-mass tail is absent for objects in our
  sample hotter than 12,000K. We interpret this as a signature that a
  significant fraction of the DZ(A) stars are convectively mixed DA
  white dwarfs that have accreted rocky material.

\item The mass distribution of DQ white dwarfs shows two distinct
  peaks, one centered at 0.55 \msun, whose carbon abundances are well
  explained by the standard carbon dredge-up scenario, and another one
  centered at $\sim$1 \msun, whose high kinematic properties are
  consistent with the idea that they represent a population of merged
  white dwarfs. We note that the location of the 0.55 \msun\ peak is
  slightly shifted towards smaller masses relative to that of DB white
  dwarfs, most probably due to some unknown opacity source in the
  ultraviolet in our DQ models.

\end{enumerate}

While traces of hydrogen are detected (or needed) in nearly all
DBZ/DZ(A) white dwarfs, its presence is extremely rare in cool DQ
stars. This indicates that the nature of these two populations of
helium-atmosphere white dwarfs are clearly distinct. The most logical
way to explain the abundance pattern and mass distributions of these
objects is to interpret hydrogen-free DB white dwarfs as progenitors
of cool DQ stars, while the other types of cool helium-atmosphere
white dwarfs, namely the DC and DZ(A) stars, would originate from both
convectively mixed DA and cooled down DB(A) white dwarfs. Within this
scenario, the rarity of both hydrogen and heavy elements in DQ white
dwarfs is also naturally explained by invoking particularly active
post-AGB phases that would eliminate practically all the remaining
hydrogen, as well as most nearby rocky objects in orbit.

The presence of hydrogen in 40\% of the hotter DQ is, however,
somewhat mysterious. We showed that this population of massive stars
have much larger space velocities than what is expected for a
relatively young star population. As originally proposed by
\citet{Dunlap2015} in the case of Hot DQ white dwarfs, this indicates
that the massive DQ stars in our sample, which are most likely
cooled-down version of the Hot DQs, would also be the result of the
merging of two C/O white dwarfs. \citet{Dufour2007Nat,Dufour2008}
proposed that the Hot DQ white dwarfs would disguise themselves as
massive DB white dwarfs until the underlying convection zone
completely dilutes the thin helium layer, effectively transforming the
object into a C/O-dominated atmosphere when the star cools to about
24,000 K. However, such massive, and most probably magnetic, DB white
dwarfs are very scarce. Another possibility is that these merger
remnants could instead hide as DA stars with very thin hydrogen
layers, or possibly as some of the massive magnetic
white dwarfs that are often found at higher effective temperatures. We
predict that spectropolarimetric or high-resolution spectroscopic
observations of massive DQ white dwarfs should reveal that most of
them are magnetic at some level.

Interestingly, we find there is a correlation between the stellar mass
and the effective temperature for these massive DQ white
dwarfs. Despite the uncertainties associated with the temperature,
abundance, and mass scales for these objects (due most probably to
some missing opacity in the ultraviolet), we believe the observed
trend to be real, and that it represents a manifestation of an
accumulation of stars at certain effective temperatures due to the
slowing of the cooling process, when the stellar core eventually
crystallizes. Future work should address the shortcomings in the
modeling of DQ white dwarfs in order to reduce the uncertainties on
the atmospheric parameters of these objects.

\acknowledgments 

This work was supported in part by NSERC (Canada) and the Fund FRQNT
(Qu\'ebec). BHD acknowledges support from the Wootton Center for
Astrophysical Plasma Properties under the United States Department of
Energy collaborative agreement DE-FOA-0001634, from the United States
Department of Energy grant under DE-SC0010623. This work has made use
of the Montreal White Dwarf Database \citep{Dufour2017}, and also the
VALD database, operated at Uppsala University, the Institute of
Astronomy RAS in Moscow, and the University of Vienna, the SIMBAD
database, operated at CDS, Strasbourg, France \citep{Simbad2000}, data
from the European Space Agency (ESA) mission {\it Gaia}
(\url{https://www.cosmos.esa.int/gaia}), processed by the {\it Gaia}
Data Processing and Analysis Consortium (DPAC,
\url{https://www.cosmos.esa.int/web/gaia/dpac/consortium}). Funding
for the DPAC has been provided by national institutions, in particular
the institutions participating in the {\it Gaia} Multilateral
Agreement \citep{Gaia2016,Gaia2018}. Funding for the Sloan Digital Sky
Survey IV has been provided by the Alfred P. Sloan Foundation, the
U.S. Department of Energy Office of Science, and the Participating
Institutions. SDSS-IV acknowledges support and resources from the
Center for High-Performance Computing at the University of Utah. The
SDSS web site is www.sdss.org. SDSS-IV is managed by the Astrophysical
Research Consortium for the Participating Institutions of the SDSS
Collaboration including the Brazilian Participation Group, the
Carnegie Institution for Science, Carnegie Mellon University, the
Chilean Participation Group, the French Participation Group,
Harvard-Smithsonian Center for Astrophysics, Instituto de
Astrof\'isica de Canarias, The Johns Hopkins University, Kavli
Institute for the Physics and Mathematics of the Universe (IPMU) /
University of Tokyo, the Korean Participation Group, Lawrence Berkeley
National Laboratory, Leibniz Institut f\"ur Astrophysik Potsdam (AIP),
Max-Planck-Institut f\"ur Astronomie (MPIA Heidelberg),
Max-Planck-Institut f\"ur Astrophysik (MPA Garching),
Max-Planck-Institut f\"ur Extraterrestrische Physik (MPE), National
Astronomical Observatories of China, New Mexico State University, New
York University, University of Notre Dame, Observat\'ario Nacional /
MCTI, The Ohio State University, Pennsylvania State University,
Shanghai Astronomical Observatory, United Kingdom Participation Group,
Universidad Nacional Aut\'onoma de M\'exico, University of Arizona,
University of Colorado Boulder, University of Oxford, University of
Portsmouth, University of Utah, University of Virginia, University of
Washington, University of Wisconsin, Vanderbilt University, and Yale
University \citep{SDSSDR14_2018}. The Pan-STARRS1 Surveys (PS1) and
the PS1 public science archive have been made possible through
contributions by the Institute for Astronomy, the University of
Hawaii, the Pan-STARRS Project Office, the Max-Planck Society and its
participating institutes, the Max Planck Institute for Astronomy,
Heidelberg and the Max Planck Institute for Extraterrestrial Physics,
Garching, The Johns Hopkins University, Durham University, the
University of Edinburgh, the Queen's University Belfast, the
Harvard-Smithsonian Center for Astrophysics, the Las Cumbres
Observatory Global Telescope Network Incorporated, the National
Central University of Taiwan, the Space Telescope Science Institute,
the National Aeronautics and Space Administration under Grant
No. NNX08AR22G issued through the Planetary Science Division of the
NASA Science Mission Directorate, the National Science Foundation
Grant No. AST-1238877, the University of Maryland, Eotvos Lorand
University (ELTE), the Los Alamos National Laboratory, and the Gordon
and Betty Moore Foundation \citep{Panstarrs2016}. \\

\textit{Softwares: } Astropy \citep{Astropy2013, Astropy2018}, Matplotlib \citep{Matplotlib}, NumPy \citep{Numpy,Numpyb}, PySpecKit \citep{Pyspeckit}.

\bibliographystyle{aasjournal}
\bibliography{references}

%
%
\small
\begin{longtable}{llllll}
\caption{Object names, spectral types, and coordinates}\\
\hline
\hline
\multicolumn{1}{c}{J Name} & \multicolumn{1}{c}{Gaia source id} & \multicolumn{1}{c}{MWDD id} & \multicolumn{1}{c}{Sp. type} & \multicolumn{1}{c}{R.A} & \multicolumn{1}{c}{Dec} \\
\hline
\endfirsthead

\caption[]{Object names, spectral types, and coordinates -- continued}\\
\hline
\hline
\multicolumn{1}{c}{J Name} & \multicolumn{1}{c}{Gaia source id} & \multicolumn{1}{c}{MWDD id} & \multicolumn{1}{c}{Sp. type} & \multicolumn{1}{c}{R.A} & \multicolumn{1}{c}{Dec} \\
\hline
\endhead

\hline
\multicolumn{6}{c}{}
\endfoot

\hline
\multicolumn{6}{c}{}
\endlastfoot

J0000$-$0026 & 2449884942327523456 & SDSS J000052.44$-$002610.5 & DQ & 0.21853 & -0.43626 \\
J0000$-$0850 & 2441060055844546816 & WD 2357$-$091 & DQ & 0.04817 & -8.83566 \\
J0002+3209 & 2873769975034246272 & SDSS J000215.65+320914.1 & DZ & 0.56522 & 32.15387 \\
J0002$-$0627 &  & SDSS J000232.35$-$062746.0 & DZ & 0.63479 & -6.46278 \\
J0003+2240 &  & SDSS J000341.93+224043.7 & DZ & 0.92471 & 22.67881 \\
J0004+0819 &  & SDSS J000418.68+081929.9 & DZ & 1.07787 & 8.32499 \\
J0004+2432 & 2849976676552398976 & SDSS J000426.95+243258.9 & DBZ & 1.11226 & 24.54971 \\
J0004+2531 & 2850199499454835968 & SDSS J000439.38+253150.6 & DZ & 1.16411 & 25.53071 \\
\end{longtable}

\newpage

%
%

\begin{center}
\small
\begin{longtable}{lllllll}
\caption{Atmospheric parameters for DZ stars. A dagger indicates that the distance is a photometric distance based on the value of \logg = 8 in the absence of a parallax.}\\
\hline
\hline
\multicolumn{1}{l}{J Name} & \multicolumn{1}{l}{\teff(K)} & \multicolumn{1}{c}{\logg} & \multicolumn{1}{c}{$M$/\msun} & \multicolumn{1}{c}{\logcahe} & \multicolumn{1}{c}{\loghhe} &
\multicolumn{1}{c}{$D$ (pc)}\\
\hline
\endfirsthead

\caption[]{Atmospheric parameters for DZ stars -- continued.}\\
\hline
\hline
\multicolumn{1}{c}{J Name} & \multicolumn{1}{c}{\teff(K)} & \multicolumn{1}{c}{\logg} & \multicolumn{1}{c}{$M$/\msun} & \multicolumn{1}{c}{\logcahe} & \multicolumn{1}{c}{\loghhe} &
\multicolumn{1}{c}{$D$ (pc)}\\
\hline
\endhead

\hline
\multicolumn{7}{c}{}
\endfoot

\hline
\multicolumn{7}{c}{}
\endlastfoot

J0002+3209&6447(121)&8.30($^{+0.23}_{-0.22}$)&0.77($^{+0.16}_{-0.14}$)&-9.00(0.18)&<-3.08&180.6(28.7)\\ 
J0002$-$0627&7306(428)&8.00&&-9.62(0.31)&<-4.05&343.6(28.3)\\ 
J0003+2240&5493(269)&8.00&&-10.27(0.22)&<-3.00&229.1(22.4)\\ 
J0004+0819&5896(234)&8.00&&-8.86(0.16)&<-3.00&337.4(27.7)\\ 
J0004+2432&10490(614)&8.26($^{+0.28}_{-0.25}$)&0.75($^{+0.18}_{-0.16}$)&-9.74(0.18)&<-5.82&287.9(49.8)\\ 
J0004+2531&12132(1141)&8.32($^{+0.81}_{-0.62}$)&0.79($^{+0.45}_{-0.36}$)&-8.27(0.35)&<-6.04&431.5(214.5)\\ 
J0005+0018&8073(208)&7.81($^{+0.14}_{-0.14}$)&0.47($^{+0.08}_{-0.07}$)&-10.05(0.09)&<-5.01&156.4(11.6)\\ 
J0005+4003&8829(427)&8.03($^{+0.09}_{-0.09}$)&0.59($^{+0.06}_{-0.05}$)&-10.42(0.04)&-4.48(0.21)&64.3(0.3)\\
\end{longtable}
\end{center}

\newpage
%
%

\begin{center}
\small
\begin{longtable}{lllllll}
\caption{Atmospheric parameters for DQ stars. A star indicate that the value of \loghhe have been fixed rather than fitted. A dagger indicates that the distance is a photometric distance based on the value of \logg = 8 in the absence of a parallax.}\\
\hline
\hline
\multicolumn{1}{c}{J Name} & \multicolumn{1}{c}{\teff(K)} & \multicolumn{1}{c}{\logg} & \multicolumn{1}{c}{$M$/\msun} & \multicolumn{1}{c}{\logche}  & \multicolumn{1}{c}{\loghhe} &
\multicolumn{1}{c}{$D$ (pc)}\\
\hline
\endfirsthead

\caption[]{Atmospheric parameters for DQ stars -- continued.}\\
\hline
\hline
\multicolumn{1}{c}{J Name} & \multicolumn{1}{c}{\teff(K)} & \multicolumn{1}{c}{\logg} & \multicolumn{1}{c}{$M$/\msun} & \multicolumn{1}{c}{\logche}  & \multicolumn{1}{c}{\loghhe} &
\multicolumn{1}{c}{$D$ (pc)} \\
\hline
\endhead

\hline
\multicolumn{7}{c}{}
\endfoot

\hline
\multicolumn{7}{c}{}
\endlastfoot

J0000$-$0026&7420(163)&7.87($^{+0.09}_{-0.08}$)&0.50($^{+0.05}_{-0.04}$)&-6.11(0.12)&&125.9(4.5)\\ 
J0000$-$0850&7853(226)&7.42($^{+0.19}_{-0.17}$)&0.30($^{+0.07}_{-0.05}$)&-5.51(0.09)&&207.3(18.4)\\ 
J0006+1800&7422(155)&7.90($^{+0.06}_{-0.06}$)&0.52($^{+0.03}_{-0.03}$)&-5.98(0.03)&&57.9(0.4)\\ 
J0007+2821&7409(195)&7.92($^{+0.15}_{-0.15}$)&0.53($^{+0.09}_{-0.08}$)&-5.89(0.10)&&184.5(15.1)\\ 
J0008+2507&10801(525)&8.10($^{+0.28}_{-0.26}$)&0.64($^{+0.18}_{-0.15}$)&-3.37(0.21)&-3.00*&311.1(53.6)\\ 
J0008$-$1034&7587(543)&7.86($^{+0.26}_{-0.24}$)&0.50($^{+0.15}_{-0.12}$)&-5.85(0.14)&&145.0(7.5)\\ 
J0015+0309&7261(217)&7.98($^{+0.21}_{-0.20}$)&0.56($^{+0.13}_{-0.11}$)&-6.03(0.12)&&199.2(23.0)\\ 
J0024$-$1107&8330(202)&7.88($^{+0.07}_{-0.07}$)&0.51($^{+0.04}_{-0.04}$)&-5.09(0.06)&&108.1(2.1)\\
\end{longtable}
\end{center}

\section{Appendix I}
\label{appendix:one}
    \clearpage

%
%
\newcounter{numpage2}
\newcounter{firstpage2}
\newcounter{lastpage2}
\setcounter{firstpage2}{2}
\setcounter{lastpage2}{171}  
\setcounter{numpage2}{\value{firstpage2}}
\stepcounter{lastpage2}

\begin{figure}[t]
\centering
\includegraphics[page=1,scale=0.65]{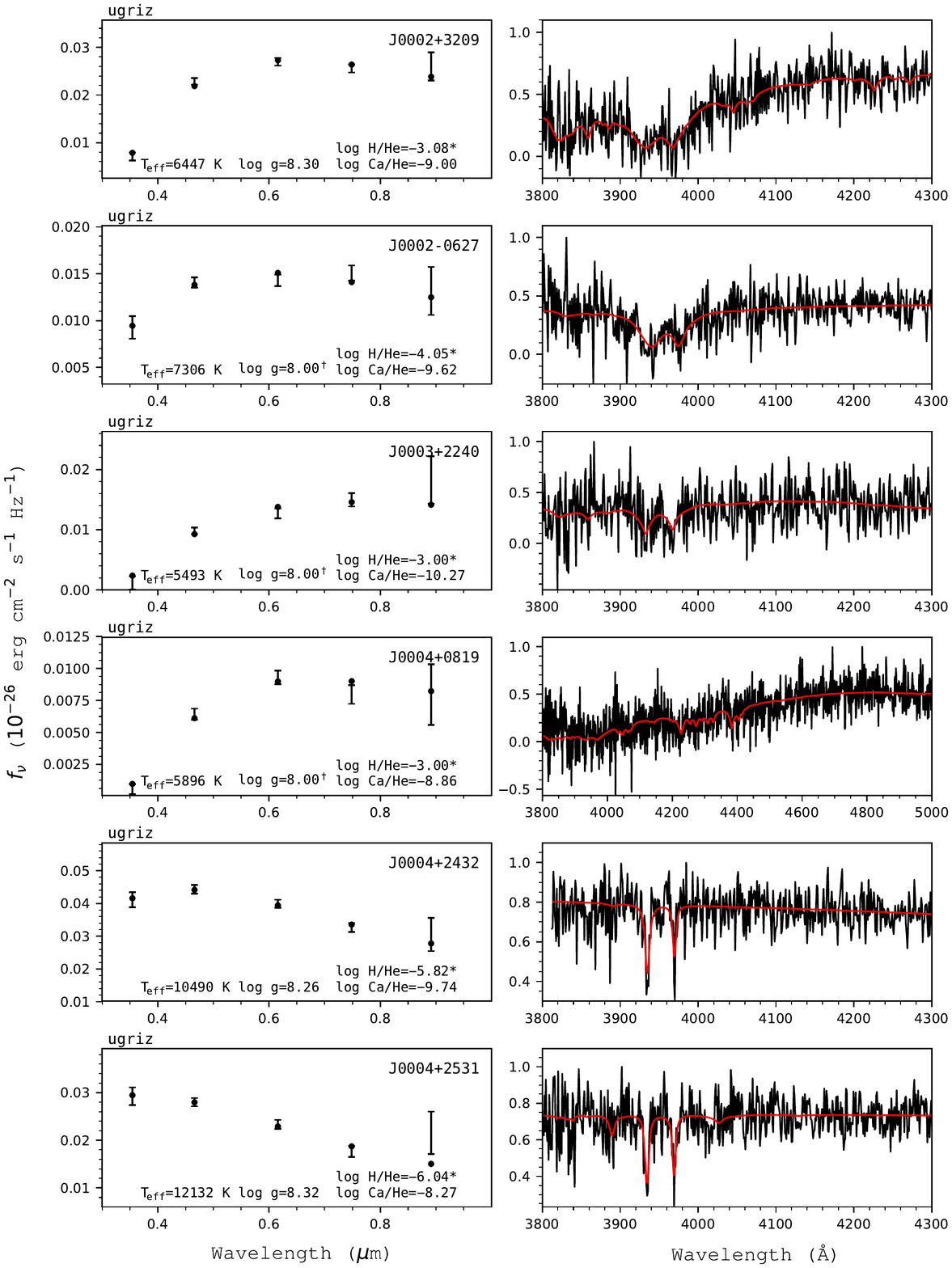}
\caption{Fits to our sample of DBZ/DZ(A) white dwarfs. Left panels: Photometric fits where error bars represent the observed data, while filled circles correspond to average model fluxes. A dagger symbol indicates that \logg is fixed at 8.0 (no parallax measurement available), while a star symbol indicates a value of \loghhe fixed at the visibility limit. Right panels: Spectroscopic fits (red) to the normalized observed spectra (black). The inset shows the fit to H$\alpha$ when present.}
\label{fig:fitsDZ}
\end{figure}	
\makeatletter	
\@whilenum{\value{numpage2}<\value{lastpage2}}\do{%
    \begin{figure}[t]
    \centering
    \includegraphics[page=\value{numpage2},scale=0.65]{dz_fitsall.pdf}\par
    \caption{Fits to the DBZ/DZ(A) white dwarfs - continued.} 
    \end{figure}
    \clearpage
\stepcounter{numpage2}}
\makeatother

%
%
\newcounter{numpage}
\newcounter{firstpage}
\newcounter{lastpage}
\setcounter{firstpage}{2}
\setcounter{lastpage}{53} 
\setcounter{numpage}{\value{firstpage}}
\stepcounter{lastpage}

\begin{figure}[t]
\centering
\includegraphics[page=1,scale=0.65]{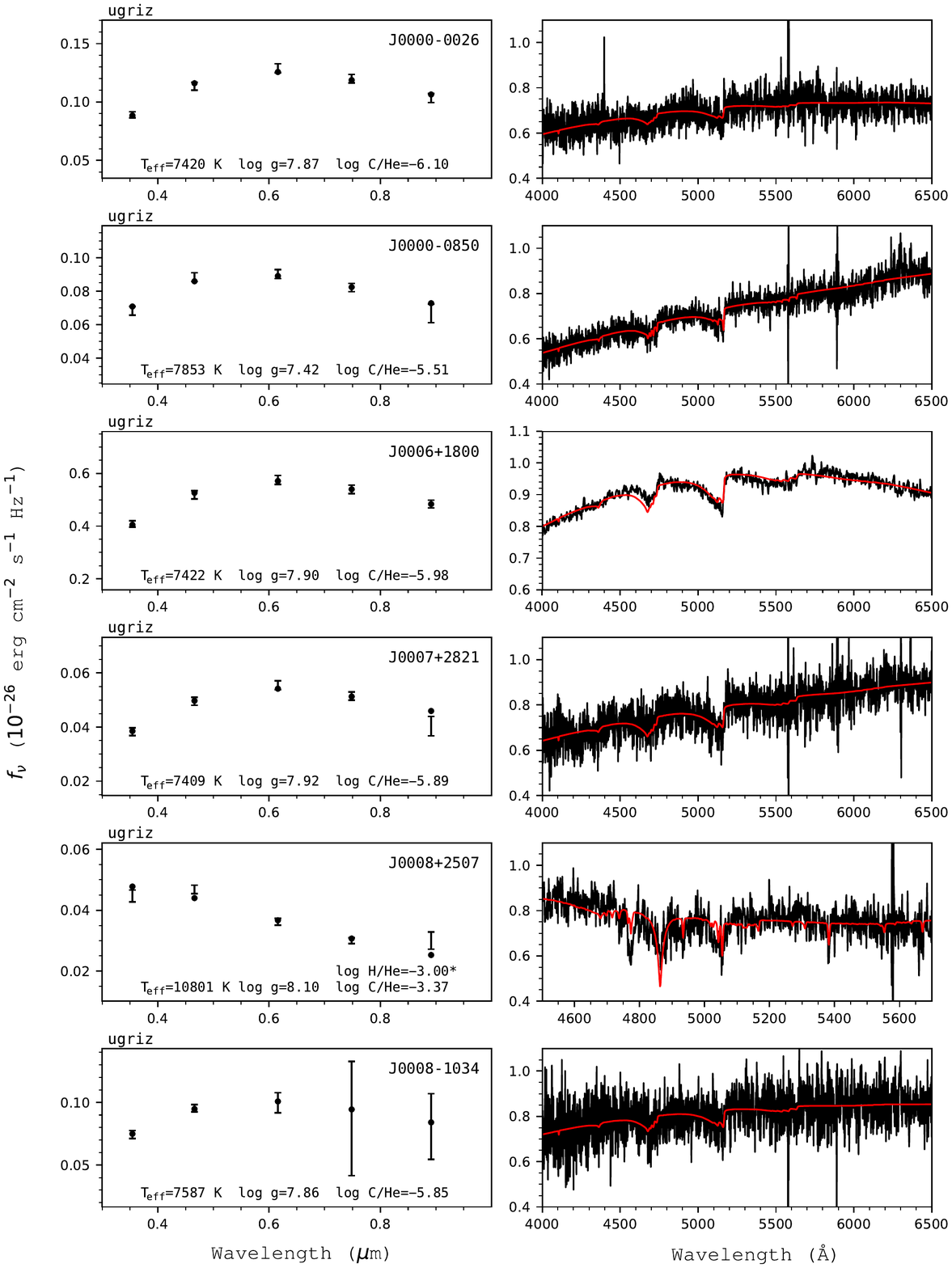}
\caption{Fits to DQ white dwarfs in our sample. In the left panels, error bars represent the observed data, while filled circles correspond to our best fit model, with the atmospheric parameters given in each panel. The photometry used in the fit is indicated at the top left of each panel. A dagger symbol indicates that the \logg value has been fixed at 8.0, when no trigonometric parallax is available. A star symbol indicates that the value of \loghhe has been fixed rather than fitted. The right panels show our spectroscopic fits (red) to the normalized observed spectra (black).}
\label{fig:fitsDQ}
\end{figure}	
\makeatletter	
\@whilenum{\value{numpage}<\value{lastpage}}\do{%
    \begin{figure}[t]
    \centering
    \includegraphics[page=\value{numpage},scale=0.65]{dq_fitsall.pdf}\par
    \caption{Fits to the DQ white dwarfs - continued.}  
    \end{figure}
    \clearpage
\stepcounter{numpage}}
\makeatother

\end{document}